\documentclass[%
 reprint,
 amsmath,amssymb,
 aps,
]{revtex4-2}

\usepackage{graphicx}
\usepackage{dcolumn}
\usepackage{bm}
\usepackage{xcolor}

\newcommand{\rf}[1]{(\ref{#1})}

\newcommand{\beq}{\begin{equation}}
\newcommand{\eeq}{\end{equation}}
\newcommand{\beqr}{\begin{eqnarray}}
\newcommand{\eeqr}{\end{eqnarray}}
\newcommand{\lb}[1]{\label{#1}}
\newcommand{\bc}{\begin{center}}
\newcommand{\ec}{\end{center}}
\newcommand{\ct}[1]{\cite{#1}}

\begin{document}

\preprint{APS/123-QED}

\title{Single photon optical bistability in a small nonlinear cavity}

\author{Igor E. Protsenko}
\email{protsenk@gmail.com}
\author{Alexander V. Uskov}
\affiliation{P.N.Lebedev Physical Institute of the RAS, Moscow 119991, Russia}

\begin{abstract}
  We investigated theoretically the bistability  in a small Fabry-Perot interferometer (FPI)  with the optical wavelength  size cavity, the nonlinear Kerr medium and only a few photons, on average, excited by the external quantum field.  Analytical expressions for  the stationary mean photon number, the bistability domain, the field and the photon number fluctuation spectra are obtained.  Multiple stationary states of the FPI cavity field with different spectra  are possible  at realistic conditions, for example, in the FPI with the  photonic crystal cavity and  the semiconductor-doped glass nonlinear medium.
\end{abstract}

\maketitle


%
\section{\label{Sec1}1. Introduction}

Recent technological progress has led to a considerable reduction of  the optical integrated circuit element size \ct{Wang:16,Liu:18} in  photonic quantum technologies (PQT) \ct{Pelucchi2022}. An  essential element of PQT  is a miniature Fabri-Perot interferometer (FPI) \ct{1159437,Zhu:21,Elshaari2020} to be  part of a variety of devices, such as the optical delay lines \ct{Zhou:18},  the
wavelength-division multiplexers  \ct{KEISER19993},  laser cavities \ct{trove.nla.gov.au/work/21304573} etc.  The bistable miniature FPI is considered an essential element for PQT and is necessary for 
 ultra-low photonic signal processing \ct{Kerckhoff:11,4054411}. FPI with a nonlinear medium has optical bistability \ct{doi:10.1080/00107518308210690, PhysRevA.19.2074,doi:10.1063/1.88632} and operates as an optical transistor \ct{PROTSENKO1994304}, where noiseless amplification is possible  \ct{PhysRevA.50.1627,PROTSENKO1994304}. 

A small FPI, with a cavity of the size of the optical wavelength, is appropriate for PQT. When the nonlinear Kerr medium is in the FPI cavity,  the cavity  refractive index  and the FPI mode frequency depend on the number of the cavity photons, leading to  {\em dispersive} optical bistability at certain conditions \ct{doi:10.1080/00107518308210690, PhysRevA.19.2074, doi:10.1063/1.88632,Bowden_book}. In particular, a non-zero detuning  between the FPI mode and the input field  frequencies is necessary for bistability    \ct{doi:10.1063/1.88632}.  The detuning reduces the number
of photons in the FPI cavity.   Thus, one expects  a small number of photons in a small FPI with the detuning, as we will see below. So it is essential  to investigate whether a  bistability is possible  with only a few, one or even less than one photon, on average, in the FPI cavity. Such investigation is complicated because the photon fluctuations cannot be neglected or considered perturbations of a small number of photons.  
The purpose of this paper is to contribute to such an investigation. 

Here we analyze, by the analytical approach, the bistability in the small FPI with the nonlinear Kerr medium, excited by the field where the quantum fluctuations are significant and not a perturbation.  
The linearized theory has previously analyzed optical bistability in the FPI with the  field quantum fluctuations
considered perturbations    \ct{Drummond_1980}. Meanwhile, the exact quantum steady-state equation of \ct{Drummond_1980}, found with the help of P-representation, did not exhibit bistability or hysteresis.

Section~\ref{section2} describes a simplified model of the nonlinear FPI, (i.e. the FPI with a nonlinear Kerr medium inside) shown in Fig.~\ref{Fig1}, with only one semitransparent mirror. We write equations of motion for such a model. 

Section~\ref{sec3} presents 
analytical formulas for the field, and the photon number fluctuation spectra of the mode of the  FPI introduced in section~\ref{section2}. The derivation of formulas of section~\ref{sec3} is given in the Appendix. Section~\ref{sec3}  generalizes results for a simple FPI in Fig.~\ref{Fig1}   to the FPI with two semitransparent mirrors shown in Fig.~\ref{fig2}. 

Section~\ref{Section_IV} describes the bistability conditions for the nonlinear FPI, excited by the external quantum field and shows the example of the FPI field spectra at the bistability.

Section~\ref{Sec_V} estimates the values of parameters necessary for  the optical bistability in the small FPI with a few photons in the cavity.

We discuss the results in the discussion section~\ref{Disc} and finalize the paper  in the conclusion   section~\ref{Conc}. 
\section{ The model and equations of motion}\label{section2}
\begin{figure}[thb]\bc
\centering
\includegraphics[width=9cm]{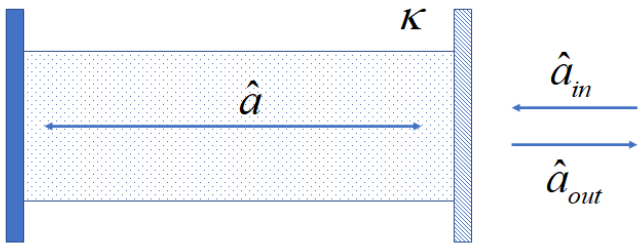}
\caption{Scheme of the nonlinear FPI with the semitransparent mirror on the right and the perfectly reflected mirror on the left. Notations are explained in the text.}
\label{Fig1}\ec
\end{figure}

%
%
To simplify the analysis, we consider, in the beginning, the  FPI with only one semitransparent and one perfectly reflecting mirror, is shown in Figure~\ref{Fig1}. Then we generalize the approach  to the FPI with two semitransparent mirrors shown in Figure~\ref{fig2}.

The  cavity of the FPI is filled with the Kerr medium, whose  refractive index depends on the cavity field intensity. The mode of the FPI cavity is excited by the quantum input field taken from a laser or a LED. The input field  Bose operator is ${{\hat{a}}_{in}}{{e}^{-i{{\omega }_{in}}t}}$;  ${{\hat{a}}_{in}}$ is the field amplitude operator, and the input field spectrum is centered on the optical carrier frequency ${{\omega }_{in}}$. The input field enters the FPI through the semitransparent mirror with the transmission rate  $\kappa$. 

We suppose that the FPI cavity length is $\lambda/2$, where $\lambda$ is the input field wavelength and assume that the main FPI cavity mode is excited; 
$\omega_{in}$  is close to the frequency ${{\omega }_{0}}$ of the center of the excited FPI mode spectrum. We neglect the  excitation of the other FPI modes.

The output field with the amplitude Bose operator ${{\hat{a}}_{out}}$ leaves the FPI through the semitransparent mirror. 
%
%
%
%
The amplitude Bose operator of the excited FPI mode is $\hat{a}$.  The input and the output field mean powers are 
%
%
$p_{in}=\left\langle \hat{a}_{in}^{+}{{{\hat{a}}}_{in}} \right\rangle$, and $p_{out}=\left\langle \hat{a}_{out}^{+}{{{\hat{a}}}_{out}} \right\rangle$, respectively.  
We denote quantum-mechanical averaging as $\left\langle ... \right\rangle $, the mean values (and the c-number coefficients)-- by letters without hats (as $a$), and operators -- by letters with hats (as $\hat{a}$).

The Hamiltonian of the FPI shown in Fig.~\ref{Fig1} written  in the interaction picture, the rotating wave approximation and with the normal ordering of Bose operators, is
\beq
H=\hbar {{\delta }_{0}}{{\hat{a}}^{+}}\hat{a}-\frac{\hbar {{\delta }_{1}}}{2}{{\hat{a}}^{+}}{{\hat{a}}^{+}}\hat{a}\hat{a}+\hat{\Gamma }, \lb{ham}
\eeq
where the detuning 
${{\delta }_{0}}={{\omega }_{0}}-{{\omega }_{in}}\ll {{\omega }_{0}},{{\omega }_{in}}$; ${{\delta }_{1}}$ is the nonlinearity coefficient of the Kerr medium, the multiplier $-1/2$ in $-\hbar {{\delta }_{1}}/2$  is introduced for convenience; 
$\hat{\Gamma }$ describes the input field coming to the FPI and the field leaving the FPI through the semitransparent mirror. Note that $\delta_1\sim \tilde{n}_2/V$, where $\tilde{n}_2$ is the Kerr medium nonlinear coefficient used in the literature \ct{Akhmanov1992}, and $V$ is the FPI cavity mode volume. A small $V$, of the size about the optical wavelength, provides a large $\delta_1$, enough for the bistability with only a few photons in the FPI cavity, as discussed in section~\ref{Disc}. 

Hamiltonian \rf{ham} leads to the Heisenberg equation of motion for $\hat{a}$
\beq
	\dot{\hat{a}}=-(i \delta_0+\kappa)\hat{a} + i\delta_1\hat{n}\hat{a}+\sqrt{2\kappa }\hat{a}_{in}, 	\lb{eqm}
\eeq
where $\hat{n}={{\hat{a}}^{+}}\hat{a}$ is the photon number operator. The terms $-\kappa \hat{a}$ and $\sqrt{2\kappa } {{{\hat{a}}}_{in}}$ describe the cavity field decay and the external field coming through the semitransparent mirror. According to the input-output theory \ct{PhysRevA.46.2766,PhysRevA.30.1386}, these terms are added to Eq.~\rf{eqm}. 

If the FPI is excited by the coherent classical  field, and quantum fluctuations of the FPI field are neglected, then the operators $\hat{a}$ and $\hat{a}_{in}$ in Eq.~\rf{eqm} must be replaced by c-number variables $a$ and ${a}_{in}$; the photon number operator $\hat{n}$ is replaced by $|a|^2$, so Eq.~\rf{eqm} turns into the classical equation has multiple stationary solutions \ct{doi:10.1080/00107518308210690, PhysRevA.19.2074, doi:10.1063/1.88632,Bowden_book}. 

The quantum equation \rf{eqm} can be solved, in principle, by the perturbation procedure: we separate in $\hat{n}$ the mean photon number $n$ and the photon number fluctuations $\delta \hat{n}$; insert $\hat{n}=n+\delta \hat{n}$ into Eq.~\rf{eqm}, neglect by $\delta \hat{n}$ and obtain 
\beq
	\dot{\hat{a}}=-(i \delta_0+\kappa)\hat{a} + i\delta_1n\hat{a}+\sqrt{2\kappa }\hat{a}_{in}. 	\lb{eqm1}
\eeq
Eq.~\rf{eqm1} is linear in operators and can be solved by the operator Fourier-transform. The mean photon number $n$  can be found from the stationary solution of Eq.~\rf{eqm1} by the procedure similar to the one in \ct{PhysRevA.59.1667,Andre:19, Protsenko_2021} used for the laser equations.

Eq.~\rf{eqm} can be approximated by Eq.~\rf{eqm1} if the photon number fluctuations are small relative to $n$. It is not true at  $n \leq 1$ when the photon number fluctuations are not small and cannot be neglected. The following sections show how to modify Eq.~\rf{eqm1} when the photon number fluctuations are not small. 
\section{Fluctuation spectra}\label{sec3}
\noindent We  represent operators in Eq.~\rf{eqm} by  Fourier-expansions
\beq
\hat{A}(t) = (2\pi)^{-1}\int_{-\infty}^{\infty}\hat{A}(\omega)e^{-i\omega t}d\omega, \lb{f_exp_gn}
\eeq
where $\hat{A}$ means $\hat{a}$,  $\hat{a}_{in}$ or $\hat{n}\hat{a}$; obtain algebraic relation for Fourier component operator $\hat{a}(\omega)$ and find a formal solution 
\beq
\hat{a}(\omega )=\frac{i{{\delta }_{1}}{{\left( \hat{n}\hat{a} \right)}_{\omega }}+\sqrt{2{\kappa}}{{{\hat{a}}}_{in}}(\omega )}{i\left( {{\delta }_{0}}-\omega  \right)+{\kappa}}. \lb{FC_sol}
\eeq
Here $\left( \hat{n}\hat{a} \right)_{\omega }$ is the Fourier-component of the operator product $\hat{n}\hat{a}$. 
Note that we use the multiplier $e^{-i\omega t}$  in Fourier expansions. In particular, 
\beq
\hat{A}^+(t) = (2\pi)^{-1}\int_{-\infty}^{\infty}(\hat{A}^+)_{\omega}e^{-i\omega t}d\omega.  \lb{f_exp_conj}
\eeq
It follows from Eqs.~\rf{f_exp_gn}, \rf{f_exp_conj} that the Fourier-component operator $(\hat{A}^+)_{\omega} = [\hat{A}(-\omega)]^+\equiv \hat{A}^+(-\omega)$.  

The stationary field spectrum  $n(\omega)$ satisfies the relation  $\left<(\hat{a}^+)_{\omega}\hat{a}(\omega')\right> = n(\omega)\delta(\omega+\omega')$. 
Using Eq.~\rf{FC_sol}, we find 
\begin{widetext}\beq
n(\omega )=\frac{\delta _{1}^{2}\left\langle {{\left( {{{\hat{a}}}^{+}}\hat{n} \right)}_{-\omega }}{{\left( \hat{n}\hat{a} \right)}_{\omega }} \right\rangle +i{{\delta }_{1}}\sqrt{2{{\kappa }}}\left[ \left\langle {\hat{a}_{in}^{+}(\omega) }{{\left( \hat{n}\hat{a} \right)}_{\omega }} \right\rangle -\left\langle {{\left( {{{\hat{a}}}^{+}}\hat{n} \right)}_{-\omega }}{{{\hat{a}}}_{in}}(\omega ) \right\rangle  \right]+2{{\kappa }}{{p}_{in}}(\omega )}{{{\left( {{\delta }_{0}}-\omega  \right)}^{2}}+\kappa^{2}}\lb{Ph_sp}
\eeq\end{widetext}
where $p_{in}(\omega )=\left<\hat{a}_{in}^+(\omega)\hat{a}_{in}(\omega)\right>$ is the input field power spectrum. In order to find $n(\omega)$ we must calculate $\left\langle {{\left( {{{\hat{a}}}^{+}}\hat{n} \right)}_{-\omega }}{{\left( \hat{n}\hat{a} \right)}_{\omega }} \right\rangle$ and $\left\langle { \hat{a}_{in}^{+}(\omega) }{{\left( \hat{n}\hat{a} \right)}_{\omega }} \right\rangle$. It is found in  Appendix that
\beq
\left\langle { \hat{a}_{in}^{+} (\omega) }{{\left( \hat{n}\hat{a} \right)}_{\omega }} \right\rangle = \frac{2n\sqrt{2\kappa }{{p}_{in}}(\omega )}{i\left( {{\delta }_{n}}-\omega  \right)+\kappa }, \lb{res1}
\eeq
where $\delta_n$ is a nonlinear detunung
\beq
{{\delta }_{n}} = \delta_0-2\delta_1n. \lb{nonlin_detun} 
\eeq
The energy conservation law for the FPI  in Fig.~\ref{Fig1} requires $p_{in}(\omega) = p_{out}(\omega)$, where $p_{out}(\omega)=\left<\hat{a}_{out}^+(\omega)\hat{a}_{out}(\omega)\right>$ is the output field power spectrum. It follows from the energy conservation law that
\beq
\left\langle {{\left( {{{\hat{a}}}^{+}}\hat{n} \right)}_{-\omega }}{{\left( \hat{n}\hat{a} \right)}_{\omega }} \right\rangle =\frac{8\kappa {{n}^{2}}{{p}_{in}}(\omega )}{{{\left( {{\delta }_{n}}-\omega  \right)}^{2}}+{{\kappa }^{2}}},\lb{nonlin_sp}
\eeq
as shown in the Appendix. Substituting the  results \rf{res1}, \rf{nonlin_sp} and $\left\langle {{\left( {{{\hat{a}}}^{+}}\hat{n} \right)}_{-\omega }}{{{\hat{a}}}_{in}}(\omega ) \right\rangle = \left\langle {{ \hat{a}_{in}^{+}(\omega) }}{{\left( \hat{n}\hat{a} \right)}_{\omega }} \right\rangle^*$ into Eq.~\rf{Ph_sp} we find the FPI cavity mode spectrum
\beq
n(\omega )=\frac{2\kappa {{p}_{in}}(\omega )}{{{\left( {{\delta }_{n}}-\omega  \right)}^{2}}+{{\kappa }^{2}}}. \lb{F_spectrum}
\eeq
In a similar way, we obtain the spectrum $(n+1)_{\omega}$ of the anti-normal ordered operator product $\hat{a}\hat{a}^+$. $(n+1)_{\omega}$ is given by Eq.~\rf{F_spectrum} with the replacement of ${p}_{in}(\omega)$ by ${p}_{in}(\omega)+1$. Thus, we find a "commutator spectrum" $[\hat{a},\hat{a}^+]_{\omega}= (n+1)_{\omega} - n(\omega)\equiv c(\omega)$
\beq
c(\omega) = \frac{2\kappa}{{{\left( {{\delta }_{n}}-\omega  \right)}^{2}}+{{\kappa }^{2}}}, \lb{com_spectrum}
\eeq
$(2\pi)^{-1}\int_{-\infty}^{\infty}[\hat{a},\hat{a}^+]_{\omega}d\omega = 1$ as it must be for the cavity mode Bose operators   \ct{PhysRevA.46.2766,PhysRevA.30.1386}.

Formulas \rf{F_spectrum}, \rf{com_spectrum} and the formula
\begin{widetext}\beq
{{\delta }^{2}}n\left( \omega  \right)=\frac{1}{2\pi }\int\limits_{-\infty }^{\infty }{n(\omega +\omega ')n(\omega ')d\omega }'+\frac{1}{4\pi }\int\limits_{-\infty }^{\infty }{\left[ n(\omega '+\omega )+n(\omega '-\omega ) \right]c(\omega ')d\omega }' \lb{fin_popfl_sp}
\eeq\end{widetext}
derived in \ct{sym15020346} let us find  the stationary field $n(\omega)$ and the photon number fluctuation ${{\delta }^{2}}n\left( \omega  \right)$ spectra of the FPI cavity mode excited by the external quantum field.   

One can see that the results \rf{F_spectrum} and \rf{com_spectrum} follow from an effective Hamiltonian 
\beq
H_{eff}=\hbar (\delta_0-2\delta_1n){{\hat{a}}^{+}}\hat{a}+\hat{\Gamma } \lb{h_eff}
\eeq
quadratic in operators $\hat{a}$ and $\hat{a}^+$. Taking into account   $\left<{{\hat{a}}^{+}}{{\hat{a}}^{+}}\hat{a}\hat{a}\right> = 2n^2$, we note that the mean  $2\hbar\delta_1n\left<{{\hat{a}}^{+}}\hat{a}\right> = 2\hbar\delta_1n^2$ of the nonlinear term in the effective Hamiltonian \rf{h_eff} is {\em twice larger} then the mean $({\hbar {{\delta }_{1}}}/{2})\left<{{\hat{a}}^{+}}{{\hat{a}}^{+}}\hat{a}\hat{a}\right> = \hbar \delta_{1}n^2$ 
 of the nonlinear term in the exact Hamiltonian \rf{ham}.

The analysis in the Appendix, leading to results \rf{F_spectrum}, \rf{com_spectrum},  takes into account the photon number fluctuations, neglected in the term $i\delta_1n\hat{a}$ in the approximate Eq.~\rf{eqm1}. The replacement $\delta_1\rightarrow2\delta_1$ in Eq.~\rf{eqm1} (equivalent to the replacement of the Hamiltonian $H$ by  $H_{eff}$) leads to the same results as the ones found in the Appendix and, therefore, enough for taking into account the photon number fluctuations related to the nonlinear term in Hamiltonian \rf{ham} -- at least for the calculations of $n$, $n(\omega)$ and $\delta^2n(\omega)$ in the stationary case. 

We use the effective Hamiltonian \rf{h_eff} to analyse of the FPI with two semi-transparent mirrors in the following  subsection.
\subsection{FPI with two semitransparent morrors}
\begin{figure}[thb]\bc
\centering
\includegraphics[width=9cm]{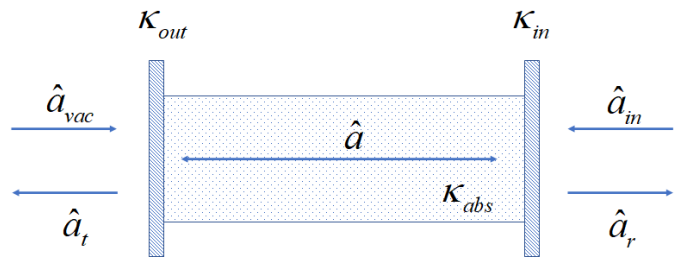}
\caption{The scheme of the nonlinear FPI with two semitransparent mirrors. $\hat{a}_{vac}$ is the vacuum field coming through the output mirror. The rest of the notations are explained in the text.}
\label{fig2}\ec
\end{figure}

Now we consider the FPI with two semitransparent mirrors, the nonlinear medium,  and the linear absorption inside the cavity. We denote $\kappa_{in}$,  ($\kappa_{out}$)  the transmission rate of the input (output) FPI mirrors;  $\kappa_{abs}$ is the rate of the linear absorption  in the FPI cavity. The field with the amplitude operator $\hat{a}_r$ is reflected from the input mirror and the field with the operator  $\hat{a}_t$ is transmitted through the FPI, as shown in Fig.~\ref{fig2}.  
%
%
%
%
The effective Hamiltonian for the FPI  in Fig.~\ref{fig2} is the same as $H_{eff}$ given by Eq.~\rf{h_eff} with the only dissipative term $\hat{\Gamma}$ different. 

There are three dissipative channels in the FPI in Fig.~\ref{fig2}: two  semitransparent mirrors and the linear absorption in the cavity. We derive Heisenberg equations of motion for $\hat{a}$ from the effective Hamiltonian \rf{h_eff}, adding dissipative terms for each channel to the equation of motion  following the input-output theory \ct{PhysRevA.46.2766,PhysRevA.30.1386}. We solve the equation of motion the same way as  Eq.~\rf{eqm}, obtaining the FPI cavity field spectrum    
\beq
n(\omega )=\frac{2\kappa_{in} {{p}_{in}}(\omega )}{{{\left( {{\delta }_{n}}-\omega  \right)}^{2}}+{\kappa_{cav}^{2}}} \lb{F_spectrum_tot}
\eeq
and the cavity field commutator spectrum (see Eq.~\rf{com_spectrum})
\beq
c(\omega) = \frac{2\kappa_{cav}}{{{\left( {{\delta }_{n}}-\omega  \right)}^{2}}+{\kappa_{cav}^{2}}}. \lb{com_spectrum_tot}
\eeq
Here 
\beq
\kappa_{cav}=\kappa_{in}+\kappa_{out}+\kappa_{abs} \lb{gammatot_1}
\eeq
is the FPI cavity mode  decay rate for all decay channels. Formulas \rf{F_spectrum_tot}, \rf{com_spectrum_tot}, together with formula \rf{fin_popfl_sp}, let us calculate the stationary mean photon number, the field and the photon number fluctuation spectra of the mode of the FPI  shown in Fig.~\ref{fig2}.  
\section{Optical bistability with a small number of photons}\label{Section_IV}
\subsection{Stationary mean values}
We re-write the photon number and the commutator spectra  \rf{F_spectrum_tot} and \rf{com_spectrum_tot} as
\beq
	n(\omega )=(\kappa_{in}/\kappa_{cav}){{{p}_{in}}(\omega )}L(\omega +\delta_n ,\kappa_{cav} ). \lb{pH_n_sp}
\eeq
and
\beq
c(\omega )=L(\omega +\delta_n ,\kappa_{cav} ). \lb{comm_sp_lor}
\eeq
Here and below,
\beq
L(\omega,\kappa) = {2\kappa}/({\omega^2+\kappa^2}) \lb{Lorenz_1}
\eeq
is a normalized Lorenz function, $(2\pi)^{-1}\int_{-\infty}^{\infty}L(\omega,\kappa)d\omega = 1$.  We take the spectrum ${{p}_{in}}(\omega )$ of the input field, with the half-width $\kappa_s$
\beq
	{{p}_{in}}(\omega )=p_{in}L(\omega,\kappa_s), \lb{pow_sp}
\eeq
where ${{p}_{in}}$ is the FPI input power in photons per second.

Using the identity $(2\pi)^{-1}\int_{-\infty}^{\infty}L(\omega+\delta,\kappa_1)L(\omega,\kappa_2) = L(\delta,\kappa_1+\kappa_2)$ and relations \rf{pH_n_sp}, \rf{pow_sp}  we calculate the mean photon number in the  FPI mode
\beq
	n=\frac{1}{2\pi }\int\limits_{-\infty }^{\infty }n(\omega )d\omega ={{p}_{eff}}L({{\delta }_n},{{\kappa }_{eff}}), \lb{n_0}
\eeq
where 
\beq
	p_{eff} = (\kappa_{in}/\kappa_{cav})p_{in}, \hspace{0.5cm} \kappa_{eff}=\kappa_s +\kappa_{cav}, \lb{gammat}
\eeq
the nonlinear detuning $\delta_n$ is determined by Eq.~\rf{nonlin_detun}.

The power of the field transmitted through the FPI is ${{p}_{out}}=2\kappa_{out} n$ or 
\beq
	{{p}_{out}}=2\kappa_{out}{p}_{eff} L({{\delta }_n},{{\kappa }_{eff}}). 	\lb{power_1}
\eeq
The result for the semi-classical case \ct{doi:10.1080/00107518308210690,PhysRevA.19.2074}  
is recovered from Eq.~\rf{power_1} at $\kappa_s=0$, so $\kappa_{eff}= \kappa_{cav}$; the multiplier 2 must be removed from the expression \rf{nonlin_detun} for the nonlinear detuning $\delta_n$.
\subsection{Bistability conditions}
We find the bistability conditions from Eq.~\rf{n_0} the way similar to the semi-classical optical bistability theory   \ct{doi:10.1080/00107518308210690} and the catastrophe theory  \ct{Poston1996,PhysRevA.19.2074}.

We introduce normalized  parameters 
\beq
Y=2p_{eff}/\kappa_{eff}, \hspace{0.2cm}  \Delta_0 = \delta_0/\kappa_{eff}, \hspace{0.2cm} \Delta_1 = 2\delta_1/\kappa_{eff},\lb{param_0}
\eeq
considering $Y$ as a function of $n$ and re-write Eq.~\rf{n_0} as 
\beq
Y=Y(n)\equiv n[1+(\Delta_0-n\Delta_1)^2].\lb{cubic_n0}
\eeq
Eq.~\rf{cubic_n0} is well-known in the semi-classical theory of  dispersive optical bistability in the cavity \ct{doi:10.1080/00107518308210690}. Parameters \rf{param_0}, however,   are different from ones in the semi-classical theory where  $2\delta_1$ and $\kappa_{eff} = \kappa_{cav}+\kappa_s$ must be  replaced, respectively, by $\delta_1$ and $\kappa_{cav}$.

\begin{figure}[thb]\bc
\centering
\includegraphics[width=7cm]{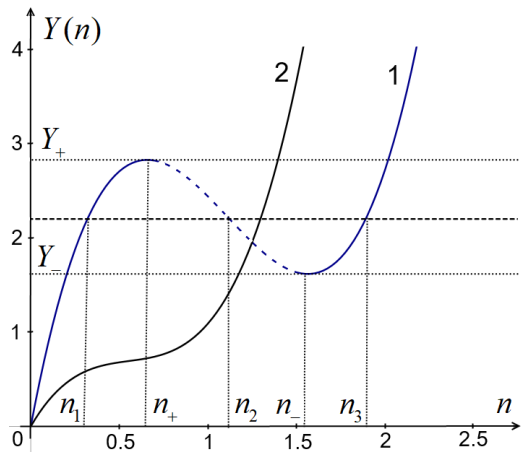}
\caption{The normalized input field power $Y$, given by Eq.~\rf{cubic_n0}, as a function of the mean photon number $n$. When the condition \rf{det_cr} is satisfied, there are three stationary $n$:  $n_1<n_2<n_3$ for some $Y$, $Y_-<Y<Y_+$, shown by the horizontal line crossing the curve 1.  The dashed part of the curve 1 corresponds to the unstable stationary solution as  $n_2$.  Eq~\rf{cubic_n0} has two stationary solutions when $Y=Y_{\pm}$. One of the solutions at $Y=Y_+$ is $n_+$, the solution at $Y=Y_-$ is $n_-$; $n_{\pm}$ is given by  Eq~\rf{n_bif}. There is only one stationary $n$ for any $Y$, when condition \rf{det_cr} is not true, as for the curve 2.}
\label{Fig3}\ec
\end{figure}

Eq.~\rf{cubic_n0} is a cubic equation for $n$ with  one, two or three real roots depending on the values of $Y$ and $\Delta_{0,1}$ \ct{Korn2000}.  We plot  $Y(n)$ in Figure~\ref{Fig3} and see, that Eq.~\rf{cubic_n0} has two   roots, when 
\beq
\partial {Y}(n) / \partial n = 0\lb{first_der}
\eeq
is satisfied together with Eq.~\rf{cubic_n0}. Solving the set of equations \rf{cubic_n0}, \rf{first_der} respectively to $n$ and $Y$ we  find the roots of Eq.~\rf{cubic_n0}
\beq
n_{\pm} = \left(2\Delta_0\pm\sqrt{\Delta_0^2-3}\right)/3\Delta_1, \lb{n_bif}
\eeq
at $Y_{\pm}= 2p_{\pm}/\kappa_{eff}$, where 
\beq
{{p}_{\pm }}=\kappa_{eff}n_{\pm}\left[1+(\Delta_0-n_{\pm}\Delta_1)^2\right]/2, \lb{p_bif}
\eeq
Eqs.~\rf{n_bif} and \rf{p_bif} defines a surface, separating regions with one and three solutions of 
Eq.~\rf{cubic_n0} in the parameter space. Three  solutions of Eq.~\rf{cubic_n0} exist at $p_-<p<p_+$, otherwise there is only one  solution, as shown in Figure~\ref{Fig3}. 

According to Eq.~\rf{n_bif} $n_{\pm}$ is real  and, therefore, the parameter region with three solutions exists, if $\Delta_0^2 \geq 3$, which means that 
\beq
|\delta_0| > \delta_{\min}=\sqrt{3}\kappa_{eff}. \lb{det_cr}
\eeq
So the absolute value of the detuning $\delta_0$  must be sufficiently large to have multiple stationary solutions. 

Following the semi-classical analysis and applying the Heisenberg correspondence principle \ct{HCP},  we suppose that the  solution $n_2$ from three  solutions $n_1<n_2<n_3$ of Eq.~\rf{cubic_n0} (see Figure~\ref{Fig3})  is unstable and two other solutions $n_{1,3}$ are stable relatively small deviations. So  Eqs.~\rf{n_bif} and \rf{p_bif} determine the borders of the FPI {\em bistability} region in the parameter  space. 

Taking $n_{\pm}$ positive we see from Eqs.~\rf{n_bif} and \rf{param_0} that  $\delta_0$ and $\delta_1$ must have the same signs for  bistability. Physically, it means a positive feedback between the mean number of photons $n$  and the nonlinear detuning $\delta_n$. Note the minus sign in Eq.~\rf{nonlin_detun} for $\delta_n$: while $n$ grows, the detuning $\delta_n$ decreases, providing, in turn, the  increase of $n$ at $\delta_0>0$. 
%
%
%
%

If we slowly decrease $p_{eff}$ from some  $p_{eff}>p_+$ or increase $p_{eff}$ from $p_{eff}<p_-$ and cross the bistability region $p_+<p<p_-$,  the transition from one stationary FPI state to another state happens at  $p_{eff}=p_{\pm}$.  
%
%
%
\begin{figure}[thb]\bc
\centering
\includegraphics[width=9cm]{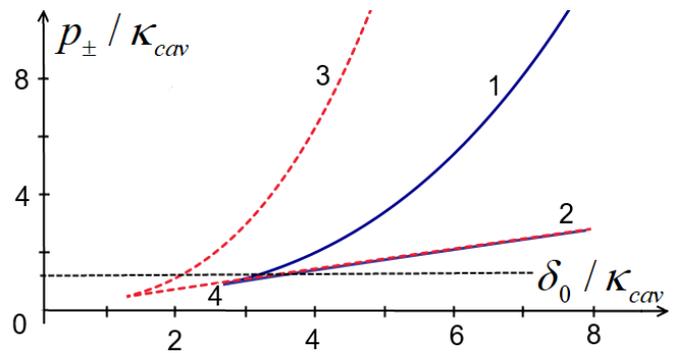}
\caption{The bistability region for the quantum FPI is between the solid curves $p_{\pm}(\delta_0)$ (curves 1 and 2; $p_+> p_-$); there is only one stationary $n$ otherwise. Curves are plotted for $\delta_1/\kappa_{cav}=1.8$, $\kappa_s/\kappa_{cav}=1$.  The dashed curves 3 and 4 restrict the bistability region for the semi-classical FPI. The dashed horizontal line marks the value $p_{eff}/\kappa_{cav} = 1.3$ taken for $n(\delta_0)$ curves 3 and 6 in Fig.~\ref{Fig5}.   }
\label{Fig4}\ec
\end{figure}
%
%
%
 
Fig.~\ref{Fig4} shows the bistability regions inside the area restricted by $p_{\pm}(\delta_0)$ solid curves for the quantum FPI with $\delta_1/\kappa_{cav} = 1.8$ and $\kappa_s/\kappa_{cav}=1$. The dashed curves in Fig.~\ref{Fig4} restrict the bistability region for the classical FPI   with $\delta_1/\kappa_{cav}=1.8$, $\kappa_s=0$ and $\delta_1$ replaced by $\delta_1/2$ in Eq.~\rf{p_bif}. The bistability region for the quantum FPI is smaller. It begins at the larger detuning than the bistability region for the classical FPI. 

%
%
\begin{figure}[thb]\bc
\centering
\includegraphics[width=9cm]{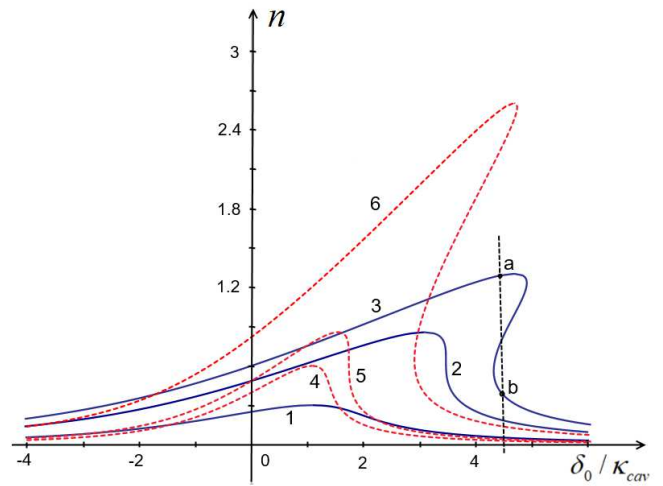}
\caption{Stationary $n(\delta_0)$ of the quantum FPI shown by the solid curves 1,2,3 for  $p_{eff}/\kappa_{cav} = 0.3$, $0.855$ and $1.3$; correspondingly, $\delta_1/\kappa_{cav}=1.8$, $\kappa_s/\kappa_{cav}=1$; and for the semi-classical FPI shown by the dashed curves 4,5,6 for $p_{eff}/\kappa_{cav} = 0.3$, $0.428$ and $1.3$. There is no bistability in curves 1 and 4 for a small $p_{eff}$; curves 2 and 5 correspond to the very beginning of the bistability region in Fig.~\ref{Fig4} at $p_+=p_-$; the bistability is for curves 3 and 6. The vertical dashed line marks $\delta_0/\kappa_{eff} = 4.4$; points $a$ and $b$ correspond to two coexisting stationary solutions with spectra shown in Fig.~\ref{Fig6}. }
\label{Fig5}\ec
\end{figure}
%
%
Fig.~\ref{Fig5} shows examples of the stationary $n(\delta_0)$ for the quantum and the classical FPI at some $p_{eff}$ values with or without the bistability in the FPI. Similar curves  are presented, for example, in \ct{Landau1976Mechanics} for the classical nonlinear oscillator. We see from Figures \ref{Fig4} and \ref{Fig5} that the bistability regions and the stationary $n$ are substantially different for the quantum and the classical FPI.
%
%
\begin{figure}[thb]\bc
\centering
\includegraphics[width=9cm]{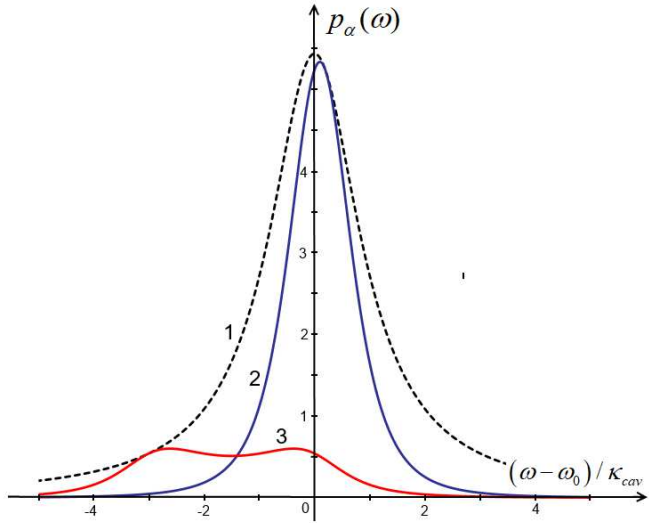}
\caption{The input field spectrum, $\alpha=in$ (the dashed curve 1) and the output field spectra $\alpha=out$ (the solid curves 2 and 3) of two stationary states of the FPI coexisting at the bistability when $\delta_0/\kappa_{cav} = 4.4$. Two stationary $n=1.3$ and $0.4$ correspond to the spectra curves 2 and 3 and points $a$ and $b$, respectively, on the $n(\delta_0)$ curve 3 in Fig.~\ref{Fig5}.  }
\label{Fig6}\ec
\end{figure}
%
%

Fig.~\ref{Fig6} shows the input field spectrum given by Eq.~\rf{pow_sp} (the dashed curve 1) and the output field spectra $p_{out}(\omega) = 2\kappa_{out}n(\omega)$ for parameters belonging to the bistability region;  $\kappa_{in}=\kappa_{out}$, $\kappa_{abs}=0$, $\delta_0/\kappa_{cav}=4.4$, $\delta_1/\kappa_{cav} = 1.8$, $\kappa_s/\kappa_{cav} =1$ and $p_{eff}/\kappa_{cav} =1.3$. The field has two stationary states  in the FPI: with $n=1.3$ and $n=0.4$; spectra 2 and 3, respectively, shown in Fig.~\ref{Fig6}, correspond to these states. Each stationary state has a specific field and photon number fluctuation spectra. The integration of $\delta^2n(\omega)$ (see Eq.~\rf{fin_popfl_sp}) over frequencies demonstrates the photon number variance $\delta^2 n = n(n+1)$   for both stationary states (with a different $n$ for each state) \ct{sym15020346}.    
\section{Parameters for  the bistability with a few photons.}\label{Sec_V}
The results found above let us estimate when bistability is possible in a small FPI with the nonlinear Kerr medium and  a few photons  inside.

According to \ct{Akhmanov1992},  the field-dependent refractive index $n_r$ of  the Kerr medium is
\beq
n_r = n_0+\tilde{n}_2I, \lb{nonlin_n}
\eeq
where $I$ is the intensity of the field, and $\tilde{n}_2$ is a nonlinearity coefficient in the refractive index, $n_0$ is the field-independent part of $n_r$. For  certainty, we  consider $\tilde{n}_2>0$ as it is in many semiconductors, such as $Si$ or $GaAs$ \ct{Akhmanov1992}. As usual, we suppose $\tilde{n}_2I\ll n_0$. 

We consider a Fabri-Perot cavity, shown in Fig.~\ref{fig2}, with Kerr medium with the nonlinear refractive index $n_r$ given by Eq.~\rf{nonlin_n}. The field of  the intensity $I$ is in the FPI main cavity mode.   
We express $n_r(I)$ in  the mean number of photons $n$ in the FPI cavity. We take $I = (n_0c/8\pi)|E|^2$ \ct{Akhmanov1997}, where $c$ is the speed of light in vacuum, and $E$ is the field amplitude. In the quantum case,    $|E|^2$  is replaced by $\left<\hat{E}^+\hat{E}\right>$, where  the field amplitude operator $\hat{E}=\sqrt{4\pi\hbar\omega_0/V}\hat{a}$, $\hat{a}$ is a Bose-operator \ct{Scully}, $V$ is the cavity mode volume,  $\omega_0$ is the carrier frequency of the FPI mode. So we re-write Eq.~\rf{nonlin_n} in  $n=\left<\hat{a}^+\hat{a}\right>$ 
\beq
n_r =  n_0+n_2n, \lb{nonlin_n_n}
\eeq
where 
\beq
n_2 = \tilde{n}_2n_0c\hbar\omega_0/2V. \lb{nonlin_n_part}
\eeq
We will estimate how large  the nonlinear coefficient $\tilde{n}_2$ of the refractive index \rf{nonlin_n} must be for  the bistability when the mean number of photons in the FPI cavity $n=1$.

The resonant frequency $\omega_m$ of the cavity mode  is \ct{Akhmanov1997}
\beq
\omega_m=\frac{\pi c m}{(n_0+n_2n)L}\approx \omega_0(1-n_2n/n_0),\lb{eigen_fr}
\eeq
where $\omega_0=\pi c m/n_0L$, $L$ is the length of the cavity, the integer $m=1$ for the FPI main cavity mode  and $n_2n\ll n_0$. We see $2\delta_1=\omega_0n_2/n_0\sim \tilde{n}_2/V$ from Eqs.~\rf{nonlin_n_part}, \rf{eigen_fr} and  \rf{nonlin_detun}.

According to conditions \rf{det_cr}, the bistability appears when the external field frequency is detuned from the cavity mode frequency at least on $\delta_0=\sqrt{3}\kappa_{eff}$ where $\kappa_{eff}$ is the sum of all linewidths and decay rates  given by Eqs.~\rf{gammat} and \rf{gammatot_1}.
We express $2\kappa_{eff}=\omega_0/Q$, where $Q$ is an effective quality factor of the main FPI cavity mode.
Expressing $\delta_1$ and $\kappa_{eff}$ in Eq.~\rf{det_cr} through $n_2$, $Q$ and other parameters, we find that the bistability is possible when $2n_2n/\sqrt{3}n_0  \geq 1/Q$ or, taking $n_2$ from Eq.~\rf{nonlin_n_part}, when
\beq
\tilde{n}_2c(\hbar\omega_0/\sqrt{3}V)nQ\geq 1. \lb{final_cond}
\eeq
Condition \rf{final_cond} with $n=1$  estimates the minimum value  $\min{\tilde{n}_2}$ of  $\tilde{n}_2$ necessary  for the bistability with only a few photons inside the FPI cavity. It must be 
\beq
\tilde{n}_2>\min{\tilde{n}_2} = \sqrt{3}V/(c\hbar\omega_0Q). \lb{min_n2}
\eeq
We take the  frequency $\omega_0$ corresponding to the wavelength $\lambda_0=1.55$~$\mu$m; the volume of  the main FPI cavity mode $V = (\lambda_0/2n_0)^3$, the linear refractive index $n_0=3.3$ as in \ct{doi:10.1063/1.5022958};    $Q=10^3$. The quality factor $Q\sim 10^3\div 10^4$ is achievable, for example,  in the photonic crystal micro-cavities  \ct{ZHANG2015374}. For such parameters, it must be  $\tilde{n}_2 > 10^{-6}~{cm^2/kWt}$  for the bistability with only a few photons in the FPI cavity. Such nonlinearity is achievable, for example, in semiconductor-doped glasses with a nonlinear response time $\sim10^{-10}\div 10^{-11}$~s \ct{Akhmanov1992}. 
\section{Discussion}\label{Disc}
We found the bistability in the stationary states of the FPI  with a small number of photons in the cavity when the FPI is excited by the quantum field and the quantum fluctuations of the field are not small. 

As a hypothesis, we assume that the upper and lower branches of the stationary $n(\delta_0)$ curve in Fig.~\ref{Fig4} correspond to the stationary states stable to small deviations from the stationary state -- as it is in the classical bistability \ct{Poston1996}.  Such a stability hypothesis must be rigorously proved  for the quantum case elsewhere in the future.

Each stationary state of the FPI  has its own fluctuations and spectra, as, for example, the field spectra shown in Fig.~\ref{Fig6}. This is the difference between the classical FPI excited by the monochromatic field  when only the stationary cavity or the output fields (with no fluctuations) can be determined.

The bistability conditions for the quantum case are different from the ones in the semi-classical case \ct{doi:10.1080/00107518308210690,PhysRevA.19.2074}. The linewidth of the input field is added to the total relaxation rate of the FPI cavity mode, as in Eq.~\rf{gammat}, and the effective nonlinear coefficient $2\delta_1$ in the nonlinear detuning \rf{nonlin_detun} is twice larger than the nonlinear coefficient $\delta_1$  in the semi-classical case. 

The transition from one  another stationary state occurs at $n=n_{\pm}$ given by Eq.~\rf{n_bif} when the stationary FPI states are on the borders of the bistability region in Fig.~\ref{Fig5}. We do not analyze the dynamics of such transitions.   Transitions between multiple classical stationary states  have been studied, for example, in \ct{Horsthemke1984}.

We have shown that the stationary mean photon number and the low-order correlations: the field, and the photon number fluctuation spectra can be found from the effective Hamiltonian \rf{h_eff}. It is shown in the Appendix that these results are a  good approximation of the exact results corresponding to the exact Hamiltonian \rf{ham} in the stationary case. We did not analyze the non-stationary dynamics (including small deviations from the stationary states) and the higher-order correlations. It will be done in the future. The low-order stationary correlations found here are enough for practical purposes in many cases \ct{Mandel1995}.

Relations \rf{gen_com}, necessary for our calculations, are held in the stationary case. With relations \rf{gen_com}, the cluster expansion approximation lets us replace the mean of the four-order operator product in the integral in \rf{three_op} by the sum \rf{cl_exp} of the binary operator products. We use the energy conservation law to determine the frequency domain's fourth-order correlations; see Eq.~\rf{res_interm} and comments  in Appendix. 

Specific field and the photon number fluctuation spectra inside and outside the cavity of the linear FPI are found in \ct{sym15020346}. The spectra of the nonlinear FPI can be obtained by replacing the detuning $\delta_0$ in \ct{sym15020346} with the nonlinear detuning $\delta_0-2\delta_1n$.

We estimate that multiple stationary solutions are possible in a small FPI cavity, of the size of the order of the optical wavelength, with a few photons and the nonlinear Kerr medium as, for example, semiconductor-doped glass. 
\section{Conclusion}\label{Conc}
We predict multiple stationary states in the theoretical model of the small Fabry-Perot interferometer (FPI) with a nonlinear Kerr medium and a few photons in the mode excited by an external quantum field. Such multiple solutions are necessary for optical bistability. The stationary mean photon number, the bistability conditions, the field and the photon number fluctuation spectra are found analytically. Estimations show that the multiple solutions appear at realistic conditions, for example, in the photonic crystal  FPI cavity of the size of the optical wavelength with a semiconductor-doped glass nonlinear medium. The results are helpful for the investigation,  construction and applications of small nonlinear elements with FPI, as optical transistors, in the photonic integrated circuits operating with quantum fields. Our treatment of the FPI with a nonlinear medium presents an example of solving the quantum nonlinear oscillator equations analytically. 

We hope the present results stimulate the experimental studies of optical bistability in a small FPI with the nonlinear Kerr medium and the quantum field.

\appendix
\section{Calculations of correlations.}
Here we find $\left\langle {{\hat{a}_{in}^{+}(\omega) }}{{\left( \hat{n}\hat{a} \right)}_{\omega }} \right\rangle$ appears in Eq.~\rf{Ph_sp}.   The operator product Fourier component is a convolution: 
\beqr
{{\left( \hat{n}\hat{a} \right)}_{\omega }}&=&(2\pi )^{-1/2}\int\limits_{-\infty }^{\infty }\hat{n}(\omega -{{\omega }_{1}})\hat{a}({{\omega }_{1}})}{d{{\omega }_{1}},\nonumber\\\hat{n}(\omega-\omega_1 )&=&(2\pi )^{-1/2}\int\limits_{-\infty }^{\infty }{{{{\hat{a}}}^{+}}({{\omega }_{2}}+\omega_1-\omega )\hat{a}({{\omega }_{2}})d{{\omega }_{2}}}\nonumber
\eeqr
Therefore, 
\[{{\left( \hat{n}\hat{a} \right)}_{\omega }}=(2\pi )^{-1/2}\int\limits_{-\infty }^{\infty }{d{{\omega }_{1}}d{{\omega }_{2}}{{{\hat{a}}}^{+}}({{\omega }_{2}}+{{\omega }_{1}}-\omega )\hat{a}({{\omega }_{2}})\hat{a}({{\omega }_{1}})}\] and
\begin{widetext}\beq
\left\langle \hat{a}_{in}^{+}(\omega ){{\left( \hat{n}\hat{a} \right)}_{\omega }} \right\rangle =\frac{1}{2\pi }\int\limits_{-\infty }^{\infty }{d{{\omega }_{1}}\int\limits_{-\infty }^{\infty }d\omega_{2} }\left\langle \hat{a}_{in}^{+}(\omega ){{{\hat{a}}}^{+}}({{\omega }_{2}}+{{\omega }_{1}}-\omega )\hat{a}({{\omega }_{2}})\hat{a}({{\omega }_{1}}) \right\rangle .\lb{three_op}
\eeq\end{widetext}
We simplify the integral \rf{three_op} using relations
\beqr
\left<\hat{a}^+(\omega')\hat{a}(\omega)\right>&=&n(\omega)\delta(\omega'-\omega), \lb{gen_com}\\ \left<\hat{a}(\omega')\hat{a}^+(\omega)\right>&=&(n+1)_{\omega}\delta(\omega'-\omega), \nonumber
\eeqr
hold in the stationary case. In Eq.~\rf{gen_com} $n(\omega)$ is the field spectrum; we denote $(n+1)_{\omega}$ the spectrum for the anti-normally ordered operator product $\hat{a}(\omega')\hat{a}^+(\omega)$. 

It follows from relations~\rf{gen_com}  that $\left<\hat{a}^+(\omega')\hat{a}(\omega)\right>=0$ and $\hat{a}^+(\omega')$ commute with $\hat{a}(\omega)$  if $\omega\neq\omega'$. Therefore, $\hat{a}^+(\omega')$ does not correlate with $\hat{a}(\omega)$ at $\omega\neq\omega'$ at the stationary case and we replace the mean value in the integral~\rf{three_op} by
\beqr
 &\left\langle \hat{a}_{in}^{+}(\omega )\hat{a}({{\omega }_{2}})\right>\left<{{{\hat{a}}}^{+}}({{\omega }_{2}}+{{\omega }_{1}}-\omega )\hat{a}({{\omega }_{1}}) \right\rangle + & \nonumber\\ &\left\langle \hat{a}_{in}^{+}(\omega )\hat{a}({{\omega }_{1}})\right>\left<{{{\hat{a}}}^{+}}({{\omega }_{2}}+{{\omega }_{1}}-\omega )\hat{a}({{\omega }_{2}}) \right\rangle & \lb{cl_exp}
\eeqr
everywhere besides  $\omega_1=\omega_2=\omega$, when arguments of operators on the right part of  Eq.~\rf{three_op} are the same. The field in the FPI is finite, so it is  reasonable to assume that the 
$\left\langle \hat{a}_{in}^{+}(\omega ){{{\hat{a}}}^{+}}(\omega )\hat{a}({{\omega }})\hat{a}({{\omega }}) \right\rangle$ does not diverge at any $\omega$. Then the single point $\omega_1=\omega_2=\omega$ gives a negligibly small contribution to the integral in Eq.~\rf{three_op}. Therefore, we can neglect the fourth-order correlation, i.e. the case $\omega_1=\omega_2=\omega$ in Eq.~\rf{three_op}, and use the expression \rf{cl_exp} instead of $\left\langle \hat{a}_{in}^{+}(\omega ){{{\hat{a}}}^{+}}({{\omega }_{2}}+{{\omega }_{1}}-\omega )\hat{a}({{\omega }_{2}})\hat{a}({{\omega }_{1}}) \right\rangle$.

We re-write Eq.~\rf{cl_exp} with the help of the first relation \rf{gen_com}
\beqr
 &\left\langle \hat{a}_{in}^{+}(\omega )\hat{a}({{\omega }_{2}})\right>n(\omega_1)\delta(\omega_2-\omega) + &\nonumber\\ & \left\langle \hat{a}_{in}^{+}(\omega )\hat{a}({{\omega }_1})\right>n(\omega_2)\delta(\omega_1-\omega), & \lb{cl_exp_1}
\eeqr
replace $\left\langle \hat{a}_{in}^{+}(\omega ){{{\hat{a}}}^{+}}({{\omega }_{2}}+{{\omega }_{1}}-\omega )\hat{a}({{\omega }_{2}})\hat{a}({{\omega }_{1}}) \right\rangle$ by the expression~\rf{cl_exp_1} in Eq.~\rf{three_op},  carry out the integration and find
\beq
\left\langle \hat{a}_{in}^{+}(\omega ){{\left( \hat{n}\hat{a} \right)}_{\omega }} \right\rangle = 2n\left\langle \hat{a}_{in}^{+}(\omega )\hat{a}({{\omega }})\right>. \lb{res_prod_1}
\eeq
The replacement of $\left\langle \hat{a}_{in}^{+}(\omega ){{{\hat{a}}}^{+}}({{\omega }_{2}}+{{\omega }_{1}}-\omega )\hat{a}({{\omega }_{2}})\hat{a}({{\omega }_{1}}) \right\rangle$ by Eq.~\rf{cl_exp} is similar to the cluster expansion method \ct{Jahnke,PhysRevA.75.013803}, where the mean of a high-order operator product is approximated employing the binary operator products. The approximation is well-known in the classical stochastic theory as a "cumulant-neglect closure" \ct{Wu_1984,10.1115/1.3173083} when the Gaussian distribution approximates the exact classical distribution, so high-order correlations became products of the second-order correlations.  In a difference between the cluster expansion and the cumulant-neglect closure approaches applied in the time domain,  we make the expansion \rf{cl_exp} in the frequency domain and use it in the integral \rf{three_op}. In a difference with the time domain, where $\left<\hat{a}^+(t)\hat{a}(t')\right>\neq 0$ if $t\neq t'$,  relations \rf{gen_com} tell that $\left<\hat{a}^+(\omega)\hat{a}(\omega')\right>= 0$ if $\omega\neq\omega'$. So the replacement of $\left\langle \hat{a}_{in}^{+}(\omega ){{{\hat{a}}}^{+}}({{\omega }_{2}}+{{\omega }_{1}}-\omega )\hat{a}({{\omega }_{2}})\hat{a}({{\omega }_{1}}) \right\rangle$ by Eq.~\rf{cl_exp} in the integral \rf{three_op} is a  good approximation at the stationary case, when relations  \rf{gen_com} are held.

The result \rf{res_prod_1} leads to some explicit expressions. We  find $\left\langle \hat{a}_{in}^{+}(\omega )\hat{a}(\omega ) \right\rangle$ deriving from Eq.~\rf{FC_sol}
\beq
\left\langle \hat{a}_{in}^{+}(\omega )\hat{a}(\omega ) \right\rangle =\frac{i{{\delta }_{1}}\left\langle \hat{a}_{in}^{+}(\omega ){{\left( \hat{n}\hat{a} \right)}_{\omega }} \right\rangle +\sqrt{2\kappa }{{p}_{in}}(\omega )}{i\left( {{\delta }_{0}}-\omega  \right)+\kappa }. \lb{ain_a}
\eeq
Substituting the result \rf{res_prod_1} into Eq.~\rf{ain_a} we obtain that 
\beq
\left\langle \hat{a}_{in}^{+}(\omega )\hat{a}(\omega ) \right\rangle =\frac{\sqrt{2\kappa }{{p}_{in}}(\omega )}{i\left( {{\delta }_{n}}-\omega  \right)+\kappa } \lb{ain_a_ex}
\eeq
where $\delta_n=\delta_0-2\delta_1n$. Inserting the result \rf{ain_a_ex} into Eq.~\rf{res_prod_1} we arrive to the result~\rf{res1}.   

Now we find $\left\langle {{\left( {{{\hat{a}}}^{+}}\hat{n} \right)}_{-\omega }}{{\left( \hat{n}\hat{a} \right)}_{\omega }} \right\rangle$ from the energy conservation law $p_{in}(\omega) = p_{out}(\omega)$, where $p_{in}(\omega) = \left<\hat{a}_{in}^+(\omega)\hat{a}_{in}(\omega)\right>$ ( $p_{out}(\omega) = \left<\hat{a}_{out}^+(\omega)\hat{a}_{out}(\omega)\right>$) is the input (output) field power spectrum. The boundary conditions at the semitransparent mirror of the FPI on Fig.~\ref{Fig1} lead to ${{\hat{a}}_{out}}=\sqrt{2\kappa }\hat{a}-{{\hat{a}}_{in}}$ and we obtain that ${{p}_{out}}(\omega )$ is
\begin{widetext}\[
\frac{2\kappa \delta _{1}^{2}\left\langle {{\left( {{{\hat{a}}}^{+}}\hat{n} \right)}_{-\omega }}{{\left( \hat{n}\hat{a} \right)}_{\omega }} \right\rangle +i\sqrt{2\kappa }{{\delta }_{1}}\left\{ 
   \left[ \kappa +i\left( {{\delta }_{0}}-\omega  \right) \right]\left\langle \hat{a}_{in}^{+}(\omega ){{\left( \hat{n}\hat{a} \right)}_{\omega }} \right\rangle -  
 \left\langle {{\left( {{{\hat{a}}}^{+}}\hat{n} \right)}_{-\omega }}{{{\hat{a}}}_{in}}(\omega ) \right\rangle \left[ \kappa -i\left( {{\delta }_{0}}-\omega  \right) \right] 
 \right\}}{{{\left( {{\delta }_{0}}-\omega  \right)}^{2}}+{{\kappa }^{2}}}+{{p}_{in}}(\omega ).
\]
The energy conservation law ${{p}_{in}}(\omega )$=${{p}_{out}}(\omega )$ requires, therefore
\beq
2\kappa \delta _{1}^{2}\left\langle {{\left( {{{\hat{a}}}^{+}}\hat{n} \right)}_{-\omega }}{{\left( \hat{n}\hat{a} \right)}_{\omega }} \right\rangle +
i\sqrt{2\kappa }{{\delta }_{1}}\left\{ 
   \left[ \kappa +i\left( {{\delta }_{0}}-\omega  \right) \right]\left\langle \hat{a}_{in}^{+}(\omega ){{\left( \hat{n}\hat{a} \right)}_{\omega }} \right\rangle - 
  \left\langle {{\left( {{{\hat{a}}}^{+}}\hat{n} \right)}_{-\omega }}{{{\hat{a}}}_{in}}(\omega ) \right\rangle \left[ \kappa -i\left( {{\delta }_{0}}-\omega  \right) \right]  \right\}=0 \lb{res_interm}\eeq\end{widetext}
Substituting the result \rf{res1} and the complex conjugated one $\left\langle {{\left( {{{\hat{a}}}^{+}}\hat{n} \right)}_{-\omega }}{{{\hat{a}}}_{in}}(\omega ) \right\rangle = \left\langle \hat{a}_{in}^{+}(\omega ){{\left( \hat{n}\hat{a} \right)}_{\omega }} \right\rangle^*$ to Eq.~\rf{res_interm} we obtain the result \rf{nonlin_sp} and  calculate the field \rf{F_spectrum} and the commutator \rf{com_spectrum_tot} spectra as explained in the main text.

\bibliography{myrefs}

\begin{thebibliography}{39}%
\makeatletter
\providecommand \@ifxundefined [1]{%
 \@ifx{#1\undefined}
}%
\providecommand \@ifnum [1]{%
 \ifnum #1\expandafter \@firstoftwo
 \else \expandafter \@secondoftwo
 \fi
}%
\providecommand \@ifx [1]{%
 \ifx #1\expandafter \@firstoftwo
 \else \expandafter \@secondoftwo
 \fi
}%
\providecommand \natexlab [1]{#1}%
\providecommand \enquote  [1]{``#1''}%
\providecommand \bibnamefont  [1]{#1}%
\providecommand \bibfnamefont [1]{#1}%
\providecommand \citenamefont [1]{#1}%
\providecommand \href@noop [0]{\@secondoftwo}%
\providecommand \href [0]{\begingroup \@sanitize@url \@href}%
\providecommand \@href[1]{\@@startlink{#1}\@@href}%
\providecommand \@@href[1]{\endgroup#1\@@endlink}%
\providecommand \@sanitize@url [0]{\catcode `\\12\catcode `\$12\catcode
  `\&12\catcode `\#12\catcode `\^12\catcode `\_12\catcode `\%12\relax}%
\providecommand \@@startlink[1]{}%
\providecommand \@@endlink[0]{}%
\providecommand \url  [0]{\begingroup\@sanitize@url \@url }%
\providecommand \@url [1]{\endgroup\@href {#1}{\urlprefix }}%
\providecommand \urlprefix  [0]{URL }%
\providecommand \Eprint [0]{\href }%
\providecommand \doibase [0]{https://doi.org/}%
\providecommand \selectlanguage [0]{\@gobble}%
\providecommand \bibinfo  [0]{\@secondoftwo}%
\providecommand \bibfield  [0]{\@secondoftwo}%
\providecommand \translation [1]{[#1]}%
\providecommand \BibitemOpen [0]{}%
\providecommand \bibitemStop [0]{}%
\providecommand \bibitemNoStop [0]{.\EOS\space}%
\providecommand \EOS [0]{\spacefactor3000\relax}%
\providecommand \BibitemShut  [1]{\csname bibitem#1\endcsname}%
\let\auto@bib@innerbib\@empty
\bibitem [{\citenamefont {Wang}\ \emph {et~al.}(2016)\citenamefont {Wang},
  \citenamefont {Sprengel}, \citenamefont {Boehm}, \citenamefont {Muneeb},
  \citenamefont {Baets}, \citenamefont {Amann},\ and\ \citenamefont
  {Roelkens}}]{Wang:16}%
  \BibitemOpen
  \bibfield  {author} {\bibinfo {author} {\bibfnamefont {R.}~\bibnamefont
  {Wang}}, \bibinfo {author} {\bibfnamefont {S.}~\bibnamefont {Sprengel}},
  \bibinfo {author} {\bibfnamefont {G.}~\bibnamefont {Boehm}}, \bibinfo
  {author} {\bibfnamefont {M.}~\bibnamefont {Muneeb}}, \bibinfo {author}
  {\bibfnamefont {R.}~\bibnamefont {Baets}}, \bibinfo {author} {\bibfnamefont
  {M.-C.}\ \bibnamefont {Amann}},\ and\ \bibinfo {author} {\bibfnamefont
  {G.}~\bibnamefont {Roelkens}},\ }\bibfield  {title} {\bibinfo {title}
  {2.3-{$\mu$}m range inp-based type-ii quantum well fabry-perot lasers
  heterogeneously integrated on a silicon photonic integrated circuit},\ }\href
  {https://doi.org/10.1364/OE.24.021081} {\bibfield  {journal} {\bibinfo
  {journal} {Opt. Express}\ }\textbf {\bibinfo {volume} {24}},\ \bibinfo
  {pages} {21081} (\bibinfo {year} {2016})}\BibitemShut {NoStop}%
\bibitem [{\citenamefont {Liu}\ \emph {et~al.}(2018)\citenamefont {Liu},
  \citenamefont {Ramirez}, \citenamefont {Vakarin}, \citenamefont {Roux},
  \citenamefont {Frigerio}, \citenamefont {Ballabio}, \citenamefont {Simola},
  \citenamefont {Alonso-Ramos}, \citenamefont {Benedikovic}, \citenamefont
  {Bouville}, \citenamefont {Vivien}, \citenamefont {Isella},\ and\
  \citenamefont {Marris-Morini}}]{Liu:18}%
  \BibitemOpen
  \bibfield  {author} {\bibinfo {author} {\bibfnamefont {Q.}~\bibnamefont
  {Liu}}, \bibinfo {author} {\bibfnamefont {J.~M.}\ \bibnamefont {Ramirez}},
  \bibinfo {author} {\bibfnamefont {V.}~\bibnamefont {Vakarin}}, \bibinfo
  {author} {\bibfnamefont {X.~L.}\ \bibnamefont {Roux}}, \bibinfo {author}
  {\bibfnamefont {J.}~\bibnamefont {Frigerio}}, \bibinfo {author}
  {\bibfnamefont {A.}~\bibnamefont {Ballabio}}, \bibinfo {author}
  {\bibfnamefont {E.~T.}\ \bibnamefont {Simola}}, \bibinfo {author}
  {\bibfnamefont {C.}~\bibnamefont {Alonso-Ramos}}, \bibinfo {author}
  {\bibfnamefont {D.}~\bibnamefont {Benedikovic}}, \bibinfo {author}
  {\bibfnamefont {D.}~\bibnamefont {Bouville}}, \bibinfo {author}
  {\bibfnamefont {L.}~\bibnamefont {Vivien}}, \bibinfo {author} {\bibfnamefont
  {G.}~\bibnamefont {Isella}},\ and\ \bibinfo {author} {\bibfnamefont
  {D.}~\bibnamefont {Marris-Morini}},\ }\bibfield  {title} {\bibinfo {title}
  {On-chip bragg grating waveguides and fabry-perot resonators for long-wave
  infrared operation up to 8.4-{$\mu$}m},\ }\href
  {https://doi.org/10.1364/OE.26.034366} {\bibfield  {journal} {\bibinfo
  {journal} {Opt. Express}\ }\textbf {\bibinfo {volume} {26}},\ \bibinfo
  {pages} {34366} (\bibinfo {year} {2018})}\BibitemShut {NoStop}%
\bibitem [{\citenamefont {Pelucchi}\ \emph {et~al.}(2022)\citenamefont
  {Pelucchi}, \citenamefont {Fagas}, \citenamefont {Aharonovich}, \citenamefont
  {Englund}, \citenamefont {Figueroa}, \citenamefont {Gong}, \citenamefont
  {Hannes}, \citenamefont {Liu}, \citenamefont {Lu}, \citenamefont {Matsuda},
  \citenamefont {Pan}, \citenamefont {Schreck}, \citenamefont {Sciarrino},
  \citenamefont {Silberhorn}, \citenamefont {Wang},\ and\ \citenamefont
  {Jons}}]{Pelucchi2022}%
  \BibitemOpen
  \bibfield  {author} {\bibinfo {author} {\bibfnamefont {E.}~\bibnamefont
  {Pelucchi}}, \bibinfo {author} {\bibfnamefont {G.}~\bibnamefont {Fagas}},
  \bibinfo {author} {\bibfnamefont {I.}~\bibnamefont {Aharonovich}}, \bibinfo
  {author} {\bibfnamefont {D.}~\bibnamefont {Englund}}, \bibinfo {author}
  {\bibfnamefont {E.}~\bibnamefont {Figueroa}}, \bibinfo {author}
  {\bibfnamefont {Q.}~\bibnamefont {Gong}}, \bibinfo {author} {\bibfnamefont
  {H.}~\bibnamefont {Hannes}}, \bibinfo {author} {\bibfnamefont
  {J.}~\bibnamefont {Liu}}, \bibinfo {author} {\bibfnamefont {C.-Y.}\
  \bibnamefont {Lu}}, \bibinfo {author} {\bibfnamefont {N.}~\bibnamefont
  {Matsuda}}, \bibinfo {author} {\bibfnamefont {J.-W.}\ \bibnamefont {Pan}},
  \bibinfo {author} {\bibfnamefont {F.}~\bibnamefont {Schreck}}, \bibinfo
  {author} {\bibfnamefont {F.}~\bibnamefont {Sciarrino}}, \bibinfo {author}
  {\bibfnamefont {C.}~\bibnamefont {Silberhorn}}, \bibinfo {author}
  {\bibfnamefont {J.}~\bibnamefont {Wang}},\ and\ \bibinfo {author}
  {\bibfnamefont {K.~D.}\ \bibnamefont {Jons}},\ }\bibfield  {title} {\bibinfo
  {title} {The potential and global outlook of integrated photonics for quantum
  technologies},\ }\href {https://doi.org/10.1038/s42254-021-00398-z}
  {\bibfield  {journal} {\bibinfo  {journal} {Nature Reviews Physics}\ }\textbf
  {\bibinfo {volume} {4}},\ \bibinfo {pages} {194} (\bibinfo {year}
  {2022})}\BibitemShut {NoStop}%
\bibitem [{\citenamefont {Chou}(2002)}]{1159437}%
  \BibitemOpen
  \bibfield  {author} {\bibinfo {author} {\bibfnamefont {S.}~\bibnamefont
  {Chou}},\ }\bibfield  {title} {\bibinfo {title} {Subwavelength optical
  elements (soes) and nanofabrications - a path to integrate optical
  communication components on a chip},\ }in\ \href@noop {} {\emph {\bibinfo
  {booktitle} {The 15th Annual Meeting of the IEEE Lasers and Electro-Optics
  Society}}},\ Vol.~\bibinfo {volume} {2}\ (\bibinfo {year} {2002})\ pp.\
  \bibinfo {pages} {574--575}\BibitemShut {NoStop}%
\bibitem [{\citenamefont {Zhu}\ \emph {et~al.}(2021)\citenamefont {Zhu},
  \citenamefont {Shao}, \citenamefont {Yu}, \citenamefont {Cheng},
  \citenamefont {Desiatov}, \citenamefont {Xin}, \citenamefont {Hu},
  \citenamefont {Holzgrafe}, \citenamefont {Ghosh}, \citenamefont
  {Shams-Ansari}, \citenamefont {Puma}, \citenamefont {Sinclair}, \citenamefont
  {Reimer}, \citenamefont {Zhang},\ and\ \citenamefont {Lon\v{c}ar}}]{Zhu:21}%
  \BibitemOpen
  \bibfield  {author} {\bibinfo {author} {\bibfnamefont {D.}~\bibnamefont
  {Zhu}}, \bibinfo {author} {\bibfnamefont {L.}~\bibnamefont {Shao}}, \bibinfo
  {author} {\bibfnamefont {M.}~\bibnamefont {Yu}}, \bibinfo {author}
  {\bibfnamefont {R.}~\bibnamefont {Cheng}}, \bibinfo {author} {\bibfnamefont
  {B.}~\bibnamefont {Desiatov}}, \bibinfo {author} {\bibfnamefont {C.~J.}\
  \bibnamefont {Xin}}, \bibinfo {author} {\bibfnamefont {Y.}~\bibnamefont
  {Hu}}, \bibinfo {author} {\bibfnamefont {J.}~\bibnamefont {Holzgrafe}},
  \bibinfo {author} {\bibfnamefont {S.}~\bibnamefont {Ghosh}}, \bibinfo
  {author} {\bibfnamefont {A.}~\bibnamefont {Shams-Ansari}}, \bibinfo {author}
  {\bibfnamefont {E.}~\bibnamefont {Puma}}, \bibinfo {author} {\bibfnamefont
  {N.}~\bibnamefont {Sinclair}}, \bibinfo {author} {\bibfnamefont
  {C.}~\bibnamefont {Reimer}}, \bibinfo {author} {\bibfnamefont
  {M.}~\bibnamefont {Zhang}},\ and\ \bibinfo {author} {\bibfnamefont
  {M.}~\bibnamefont {Lon\v{c}ar}},\ }\bibfield  {title} {\bibinfo {title}
  {Integrated photonics on thin-film lithium niobate},\ }\href
  {https://doi.org/10.1364/AOP.411024} {\bibfield  {journal} {\bibinfo
  {journal} {Adv. Opt. Photon.}\ }\textbf {\bibinfo {volume} {13}},\ \bibinfo
  {pages} {242} (\bibinfo {year} {2021})}\BibitemShut {NoStop}%
\bibitem [{\citenamefont {Elshaari}\ \emph {et~al.}(2020)\citenamefont
  {Elshaari}, \citenamefont {Pernice}, \citenamefont {Srinivasan},
  \citenamefont {Benson},\ and\ \citenamefont {Zwiller}}]{Elshaari2020}%
  \BibitemOpen
  \bibfield  {author} {\bibinfo {author} {\bibfnamefont {A.~W.}\ \bibnamefont
  {Elshaari}}, \bibinfo {author} {\bibfnamefont {W.}~\bibnamefont {Pernice}},
  \bibinfo {author} {\bibfnamefont {K.}~\bibnamefont {Srinivasan}}, \bibinfo
  {author} {\bibfnamefont {O.}~\bibnamefont {Benson}},\ and\ \bibinfo {author}
  {\bibfnamefont {V.}~\bibnamefont {Zwiller}},\ }\bibfield  {title} {\bibinfo
  {title} {Hybrid integrated quantum photonic circuits},\ }\href
  {https://doi.org/10.1038/s41566-020-0609-x} {\bibfield  {journal} {\bibinfo
  {journal} {Nature Photonics}\ }\textbf {\bibinfo {volume} {14}},\ \bibinfo
  {pages} {285} (\bibinfo {year} {2020})}\BibitemShut {NoStop}%
\bibitem [{\citenamefont {Zhou}\ \emph {et~al.}(2018)\citenamefont {Zhou},
  \citenamefont {Wang}, \citenamefont {Lu},\ and\ \citenamefont
  {Chen}}]{Zhou:18}%
  \BibitemOpen
  \bibfield  {author} {\bibinfo {author} {\bibfnamefont {L.}~\bibnamefont
  {Zhou}}, \bibinfo {author} {\bibfnamefont {X.}~\bibnamefont {Wang}}, \bibinfo
  {author} {\bibfnamefont {L.}~\bibnamefont {Lu}},\ and\ \bibinfo {author}
  {\bibfnamefont {J.}~\bibnamefont {Chen}},\ }\bibfield  {title} {\bibinfo
  {title} {Integrated optical delay lines: a review and perspective
  (invited)},\ }\href
  {https://opg.optica.org/col/abstract.cfm?URI=col-16-10-101301} {\bibfield
  {journal} {\bibinfo  {journal} {Chin. Opt. Lett.}\ }\textbf {\bibinfo
  {volume} {16}},\ \bibinfo {pages} {101301} (\bibinfo {year}
  {2018})}\BibitemShut {NoStop}%
\bibitem [{\citenamefont {Keiser}(1999)}]{KEISER19993}%
  \BibitemOpen
  \bibfield  {author} {\bibinfo {author} {\bibfnamefont {G.~E.}\ \bibnamefont
  {Keiser}},\ }\bibfield  {title} {\bibinfo {title} {A review of wdm technology
  and applications},\ }\href
  {https://doi.org/https://doi.org/10.1006/ofte.1998.0275} {\bibfield
  {journal} {\bibinfo  {journal} {Optical Fiber Technology}\ }\textbf {\bibinfo
  {volume} {5}},\ \bibinfo {pages} {3} (\bibinfo {year} {1999})}\BibitemShut
  {NoStop}%
\bibitem [{\citenamefont {Sargent}\ \emph {et~al.}(1974)\citenamefont
  {Sargent}, \citenamefont {Scully},\ and\ \citenamefont
  {Lamb}}]{trove.nla.gov.au/work/21304573}%
  \BibitemOpen
  \bibfield  {author} {\bibinfo {author} {\bibfnamefont {M.}~\bibnamefont
  {Sargent}}, \bibinfo {author} {\bibfnamefont {M.~O.}\ \bibnamefont
  {Scully}},\ and\ \bibinfo {author} {\bibfnamefont {W.~E.}\ \bibnamefont
  {Lamb}},\ }\href@noop {} {\emph {\bibinfo {title} {Laser Physics}}}\
  (\bibinfo  {publisher} {London : Addison-Wesley},\ \bibinfo {year}
  {1974})\BibitemShut {NoStop}%
\bibitem [{\citenamefont {Kerckhoff}\ \emph {et~al.}(2011)\citenamefont
  {Kerckhoff}, \citenamefont {Armen},\ and\ \citenamefont
  {Mabuchi}}]{Kerckhoff:11}%
  \BibitemOpen
  \bibfield  {author} {\bibinfo {author} {\bibfnamefont {J.}~\bibnamefont
  {Kerckhoff}}, \bibinfo {author} {\bibfnamefont {M.~A.}\ \bibnamefont
  {Armen}},\ and\ \bibinfo {author} {\bibfnamefont {H.}~\bibnamefont
  {Mabuchi}},\ }\bibfield  {title} {\bibinfo {title} {Remnants of semiclassical
  bistability in the few-photon regime of cavity qed},\ }\href
  {https://doi.org/10.1364/OE.19.024468} {\bibfield  {journal} {\bibinfo
  {journal} {Opt. Express}\ }\textbf {\bibinfo {volume} {19}},\ \bibinfo
  {pages} {24468} (\bibinfo {year} {2011})}\BibitemShut {NoStop}%
\bibitem [{\citenamefont {Kim}\ \emph {et~al.}(2006)\citenamefont {Kim},
  \citenamefont {Hwang},\ and\ \citenamefont {Lee}}]{4054411}%
  \BibitemOpen
  \bibfield  {author} {\bibinfo {author} {\bibfnamefont {M.-k.}\ \bibnamefont
  {Kim}}, \bibinfo {author} {\bibfnamefont {I.-k.}\ \bibnamefont {Hwang}},\
  and\ \bibinfo {author} {\bibfnamefont {Y.-h.}\ \bibnamefont {Lee}},\
  }\bibfield  {title} {\bibinfo {title} {All-optical bistability in photonic
  crystal resonators based on ingaasp quantum-wells},\ }in\ \href
  {https://doi.org/10.1109/LEOS.2006.278976} {\emph {\bibinfo {booktitle} {LEOS
  2006 - 19th Annual Meeting of the IEEE Lasers and Electro-Optics Society}}}\
  (\bibinfo {year} {2006})\ pp.\ \bibinfo {pages} {769--770}\BibitemShut
  {NoStop}%
\bibitem [{\citenamefont {Lugiato}(1983)}]{doi:10.1080/00107518308210690}%
  \BibitemOpen
  \bibfield  {author} {\bibinfo {author} {\bibfnamefont {L.~A.}\ \bibnamefont
  {Lugiato}},\ }\bibfield  {title} {\bibinfo {title} {Optical bistability},\
  }\href@noop {} {\bibfield  {journal} {\bibinfo  {journal} {Contemporary
  Physics}\ }\textbf {\bibinfo {volume} {24}},\ \bibinfo {pages} {333}
  (\bibinfo {year} {1983})}\BibitemShut {NoStop}%
\bibitem [{\citenamefont {Agrawal}\ and\ \citenamefont
  {Carmichael}(1979)}]{PhysRevA.19.2074}%
  \BibitemOpen
  \bibfield  {author} {\bibinfo {author} {\bibfnamefont {G.~P.}\ \bibnamefont
  {Agrawal}}\ and\ \bibinfo {author} {\bibfnamefont {H.~J.}\ \bibnamefont
  {Carmichael}},\ }\bibfield  {title} {\bibinfo {title} {Optical bistability
  through nonlinear dispersion and absorption},\ }\href
  {https://doi.org/10.1103/PhysRevA.19.2074} {\bibfield  {journal} {\bibinfo
  {journal} {Phys. Rev. A}\ }\textbf {\bibinfo {volume} {19}},\ \bibinfo
  {pages} {2074} (\bibinfo {year} {1979})}\BibitemShut {NoStop}%
\bibitem [{\citenamefont {Felber}\ and\ \citenamefont
  {Marburger}(1976)}]{doi:10.1063/1.88632}%
  \BibitemOpen
  \bibfield  {author} {\bibinfo {author} {\bibfnamefont {F.~S.}\ \bibnamefont
  {Felber}}\ and\ \bibinfo {author} {\bibfnamefont {J.~H.}\ \bibnamefont
  {Marburger}},\ }\bibfield  {title} {\bibinfo {title} {Theory of nonresonant
  multistable optical devices},\ }\href@noop {} {\bibfield  {journal} {\bibinfo
   {journal} {Applied Physics Letters}\ }\textbf {\bibinfo {volume} {28}},\
  \bibinfo {pages} {731} (\bibinfo {year} {1976})}\BibitemShut {NoStop}%
\bibitem [{\citenamefont {Protsenko}\ and\ \citenamefont
  {Lugiato}(1994)}]{PROTSENKO1994304}%
  \BibitemOpen
  \bibfield  {author} {\bibinfo {author} {\bibfnamefont {I.}~\bibnamefont
  {Protsenko}}\ and\ \bibinfo {author} {\bibfnamefont {L.}~\bibnamefont
  {Lugiato}},\ }\bibfield  {title} {\bibinfo {title} {Noiseless amplification
  in the optical transistor},\ }\href
  {https://doi.org/https://doi.org/10.1016/0030-4018(94)90697-1} {\bibfield
  {journal} {\bibinfo  {journal} {Optics Communications}\ }\textbf {\bibinfo
  {volume} {109}},\ \bibinfo {pages} {304} (\bibinfo {year}
  {1994})}\BibitemShut {NoStop}%
\bibitem [{\citenamefont {Protsenko}\ \emph {et~al.}(1994)\citenamefont
  {Protsenko}, \citenamefont {Lugiato},\ and\ \citenamefont
  {Fabre}}]{PhysRevA.50.1627}%
  \BibitemOpen
  \bibfield  {author} {\bibinfo {author} {\bibfnamefont {I.~E.}\ \bibnamefont
  {Protsenko}}, \bibinfo {author} {\bibfnamefont {L.~A.}\ \bibnamefont
  {Lugiato}},\ and\ \bibinfo {author} {\bibfnamefont {C.}~\bibnamefont
  {Fabre}},\ }\bibfield  {title} {\bibinfo {title} {Spectral analysis of the
  degenerate optical parametric oscillator as a noiseless amplifier},\ }\href
  {https://doi.org/10.1103/PhysRevA.50.1627} {\bibfield  {journal} {\bibinfo
  {journal} {Phys. Rev. A}\ }\textbf {\bibinfo {volume} {50}},\ \bibinfo
  {pages} {1627} (\bibinfo {year} {1994})}\BibitemShut {NoStop}%
\bibitem [{\citenamefont {Bowden}\ \emph {et~al.}(1981)\citenamefont {Bowden},
  \citenamefont {Ciftan},\ and\ \citenamefont {Robl}}]{Bowden_book}%
  \BibitemOpen
  \bibfield  {author} {\bibinfo {author} {\bibfnamefont {C.~M.}\ \bibnamefont
  {Bowden}}, \bibinfo {author} {\bibfnamefont {M.}~\bibnamefont {Ciftan}},\
  and\ \bibinfo {author} {\bibfnamefont {H.~R.}\ \bibnamefont {Robl}},\
  }\href@noop {} {\emph {\bibinfo {title} {Optical bistability}}}\ (\bibinfo
  {publisher} {Plenum Press},\ \bibinfo {year} {1981})\ p.\ \bibinfo {pages}
  {614}\BibitemShut {NoStop}%
\bibitem [{\citenamefont {Drummond}\ and\ \citenamefont
  {Walls}(1980)}]{Drummond_1980}%
  \BibitemOpen
  \bibfield  {author} {\bibinfo {author} {\bibfnamefont {P.~D.}\ \bibnamefont
  {Drummond}}\ and\ \bibinfo {author} {\bibfnamefont {D.~F.}\ \bibnamefont
  {Walls}},\ }\bibfield  {title} {\bibinfo {title} {Quantum theory of optical
  bistability. i. nonlinear polarisability model},\ }\href
  {https://doi.org/10.1088/0305-4470/13/2/034} {\bibfield  {journal} {\bibinfo
  {journal} {Journal of Physics A: Mathematical and General}\ }\textbf
  {\bibinfo {volume} {13}},\ \bibinfo {pages} {725} (\bibinfo {year}
  {1980})}\BibitemShut {NoStop}%
\bibitem [{\citenamefont {Akhmanov}\ \emph {et~al.}(1992)\citenamefont
  {Akhmanov}, \citenamefont {Vysloukh},\ and\ \citenamefont
  {Chirkin}}]{Akhmanov1992}%
  \BibitemOpen
  \bibfield  {author} {\bibinfo {author} {\bibfnamefont {S.~A.}\ \bibnamefont
  {Akhmanov}}, \bibinfo {author} {\bibfnamefont {V.~A.}\ \bibnamefont
  {Vysloukh}},\ and\ \bibinfo {author} {\bibfnamefont {A.~S.}\ \bibnamefont
  {Chirkin}},\ }\href@noop {} {\emph {\bibinfo {title} {Optics of femtosecond
  laser pulses}}}\ (\bibinfo  {publisher} {American Institute of Physics},\
  \bibinfo {year} {1992})\ p.\ \bibinfo {pages} {366}\BibitemShut {NoStop}%
\bibitem [{\citenamefont {Courty}\ and\ \citenamefont
  {Reynaud}(1992)}]{PhysRevA.46.2766}%
  \BibitemOpen
  \bibfield  {author} {\bibinfo {author} {\bibfnamefont {J.-M.}\ \bibnamefont
  {Courty}}\ and\ \bibinfo {author} {\bibfnamefont {S.}~\bibnamefont
  {Reynaud}},\ }\bibfield  {title} {\bibinfo {title} {Generalized linear
  input-output theory for quantum fluctuations},\ }\href
  {https://doi.org/10.1103/PhysRevA.46.2766} {\bibfield  {journal} {\bibinfo
  {journal} {Phys. Rev. A}\ }\textbf {\bibinfo {volume} {46}},\ \bibinfo
  {pages} {2766} (\bibinfo {year} {1992})}\BibitemShut {NoStop}%
\bibitem [{\citenamefont {Collett}\ and\ \citenamefont
  {Gardiner}(1984)}]{PhysRevA.30.1386}%
  \BibitemOpen
  \bibfield  {author} {\bibinfo {author} {\bibfnamefont {M.~J.}\ \bibnamefont
  {Collett}}\ and\ \bibinfo {author} {\bibfnamefont {C.~W.}\ \bibnamefont
  {Gardiner}},\ }\bibfield  {title} {\bibinfo {title} {Squeezing of intracavity
  and traveling-wave light fields produced in parametric amplification},\
  }\href {https://doi.org/10.1103/PhysRevA.30.1386} {\bibfield  {journal}
  {\bibinfo  {journal} {Phys. Rev. A}\ }\textbf {\bibinfo {volume} {30}},\
  \bibinfo {pages} {1386} (\bibinfo {year} {1984})}\BibitemShut {NoStop}%
\bibitem [{\citenamefont {Protsenko}\ \emph {et~al.}(1999)\citenamefont
  {Protsenko}, \citenamefont {Domokos}, \citenamefont {Lef\`evre-Seguin},
  \citenamefont {Hare}, \citenamefont {Raimond},\ and\ \citenamefont
  {Davidovich}}]{PhysRevA.59.1667}%
  \BibitemOpen
  \bibfield  {author} {\bibinfo {author} {\bibfnamefont {I.}~\bibnamefont
  {Protsenko}}, \bibinfo {author} {\bibfnamefont {P.}~\bibnamefont {Domokos}},
  \bibinfo {author} {\bibfnamefont {V.}~\bibnamefont {Lef\`evre-Seguin}},
  \bibinfo {author} {\bibfnamefont {J.}~\bibnamefont {Hare}}, \bibinfo {author}
  {\bibfnamefont {J.~M.}\ \bibnamefont {Raimond}},\ and\ \bibinfo {author}
  {\bibfnamefont {L.}~\bibnamefont {Davidovich}},\ }\bibfield  {title}
  {\bibinfo {title} {Quantum theory of a thresholdless laser},\ }\href
  {https://doi.org/10.1103/PhysRevA.59.1667} {\bibfield  {journal} {\bibinfo
  {journal} {Phys. Rev. A}\ }\textbf {\bibinfo {volume} {59}},\ \bibinfo
  {pages} {1667} (\bibinfo {year} {1999})}\BibitemShut {NoStop}%
\bibitem [{\citenamefont {Andr\'{e}}\ \emph {et~al.}(2019)\citenamefont
  {Andr\'{e}}, \citenamefont {Protsenko}, \citenamefont {Uskov}, \citenamefont
  {M{\o}rk},\ and\ \citenamefont {Wubs}}]{Andre:19}%
  \BibitemOpen
  \bibfield  {author} {\bibinfo {author} {\bibfnamefont {E.~C.}\ \bibnamefont
  {Andr\'{e}}}, \bibinfo {author} {\bibfnamefont {I.~E.}\ \bibnamefont
  {Protsenko}}, \bibinfo {author} {\bibfnamefont {A.~V.}\ \bibnamefont
  {Uskov}}, \bibinfo {author} {\bibfnamefont {J.}~\bibnamefont {M{\o}rk}},\
  and\ \bibinfo {author} {\bibfnamefont {M.}~\bibnamefont {Wubs}},\ }\bibfield
  {title} {\bibinfo {title} {On collective {R}abi splitting in nanolasers and
  nano-{LED}s},\ }\href {https://doi.org/10.1364/OL.44.001415} {\bibfield
  {journal} {\bibinfo  {journal} {Opt. Lett.}\ }\textbf {\bibinfo {volume}
  {44}},\ \bibinfo {pages} {1415} (\bibinfo {year} {2019})}\BibitemShut
  {NoStop}%
\bibitem [{\citenamefont {Protsenko}\ \emph {et~al.}(2021)\citenamefont
  {Protsenko}, \citenamefont {Uskov}, \citenamefont {Andr{\'{e}}},
  \citenamefont {M{\o}rk},\ and\ \citenamefont {Wubs}}]{Protsenko_2021}%
  \BibitemOpen
  \bibfield  {author} {\bibinfo {author} {\bibfnamefont {I.~E.}\ \bibnamefont
  {Protsenko}}, \bibinfo {author} {\bibfnamefont {A.~V.}\ \bibnamefont
  {Uskov}}, \bibinfo {author} {\bibfnamefont {E.~C.}\ \bibnamefont
  {Andr{\'{e}}}}, \bibinfo {author} {\bibfnamefont {J.}~\bibnamefont
  {M{\o}rk}},\ and\ \bibinfo {author} {\bibfnamefont {M.}~\bibnamefont
  {Wubs}},\ }\bibfield  {title} {\bibinfo {title} {Quantum langevin approach
  for superradiant nanolasers},\ }\href
  {https://doi.org/10.1088/1367-2630/abfd4c} {\bibfield  {journal} {\bibinfo
  {journal} {New Journal of Physics}\ }\textbf {\bibinfo {volume} {23}},\
  \bibinfo {pages} {063010} (\bibinfo {year} {2021})}\BibitemShut {NoStop}%
\bibitem [{\citenamefont {Protsenko}\ and\ \citenamefont
  {Uskov}(2023)}]{sym15020346}%
  \BibitemOpen
  \bibfield  {author} {\bibinfo {author} {\bibfnamefont {I.~E.}\ \bibnamefont
  {Protsenko}}\ and\ \bibinfo {author} {\bibfnamefont {A.~V.}\ \bibnamefont
  {Uskov}},\ }\bibfield  {title} {\bibinfo {title} {Quantum fluctuations in the
  small fabry-perot interferometer},\ }\bibfield  {journal} {\bibinfo
  {journal} {Symmetry}\ }\textbf {\bibinfo {volume} {15}},\ \href
  {https://doi.org/10.3390/sym15020346} {10.3390/sym15020346} (\bibinfo {year}
  {2023})\BibitemShut {NoStop}%
\bibitem [{\citenamefont {Poston}\ and\ \citenamefont
  {Stewart}(1996)}]{Poston1996}%
  \BibitemOpen
  \bibfield  {author} {\bibinfo {author} {\bibfnamefont {T.}~\bibnamefont
  {Poston}}\ and\ \bibinfo {author} {\bibfnamefont {I.}~\bibnamefont
  {Stewart}},\ }\href@noop {} {\emph {\bibinfo {title} {Catastrophe theory and
  its applications}}}\ (\bibinfo  {publisher} {Dover Publications},\ \bibinfo
  {year} {1996})\ p.\ \bibinfo {pages} {491}\BibitemShut {NoStop}%
\bibitem [{\citenamefont {Korn}(2000)}]{Korn2000}%
  \BibitemOpen
  \bibfield  {author} {\bibinfo {author} {\bibfnamefont {G.~A.}\ \bibnamefont
  {Korn}},\ }\href@noop {} {\emph {\bibinfo {title} {Mathematical handbook for
  scientists and engineers}}}\ (\bibinfo  {publisher} {Dover Publications},\
  \bibinfo {year} {2000})\ p.\ \bibinfo {pages} {1130}\BibitemShut {NoStop}%
\bibitem [{\citenamefont {Dahl}(2002)}]{HCP}%
  \BibitemOpen
  \bibfield  {author} {\bibinfo {author} {\bibfnamefont {J.~P.}\ \bibnamefont
  {Dahl}},\ }\bibinfo {title} {The bohr-heisenberg correspondence principle
  viewed from phase space},\ in\ \href@noop {} {\emph {\bibinfo {booktitle}
  {100 Years Werner Heisenberg}}}\ (\bibinfo  {publisher} {John Wiley \& Sons,
  Ltd},\ \bibinfo {year} {2002})\ pp.\ \bibinfo {pages} {201--206}\BibitemShut
  {NoStop}%
\bibitem [{\citenamefont {Landau}\ and\ \citenamefont
  {Lifshitz}(1976)}]{Landau1976Mechanics}%
  \BibitemOpen
  \bibfield  {author} {\bibinfo {author} {\bibfnamefont {L.~D.}\ \bibnamefont
  {Landau}}\ and\ \bibinfo {author} {\bibfnamefont {E.~M.}\ \bibnamefont
  {Lifshitz}},\ }\href {http://www.worldcat.org/isbn/0750628960} {\emph
  {\bibinfo {title} {Mechanics, Third Edition: Volume 1 (Course of Theoretical
  Physics)}}},\ \bibinfo {edition} {3rd}\ ed.\ (\bibinfo  {publisher}
  {Butterworth-Heinemann},\ \bibinfo {year} {1976})\BibitemShut {NoStop}%
\bibitem [{\citenamefont {Akhmanov}\ and\ \citenamefont
  {Nikitin}(1997)}]{Akhmanov1997}%
  \BibitemOpen
  \bibfield  {author} {\bibinfo {author} {\bibfnamefont {S.~A.}\ \bibnamefont
  {Akhmanov}}\ and\ \bibinfo {author} {\bibfnamefont {S.~Y.}\ \bibnamefont
  {Nikitin}},\ }\href@noop {} {\emph {\bibinfo {title} {Physical optics}}}\
  (\bibinfo  {publisher} {Clarendon Press},\ \bibinfo {year} {1997})\ p.\
  \bibinfo {pages} {488}\BibitemShut {NoStop}%
\bibitem [{\citenamefont {Scully}(1997)}]{Scully}%
  \BibitemOpen
  \bibfield  {author} {\bibinfo {author} {\bibfnamefont {M.~S.}\ \bibnamefont
  {Scully}, \bibfnamefont {M.~O.~Zubairy}},\ }\href@noop {} {\emph {\bibinfo
  {title} {Quantum {O}ptics}}}\ (\bibinfo  {publisher} {Cambridge University
  Press},\ \bibinfo {year} {1997})\BibitemShut {NoStop}%
\bibitem [{\citenamefont {{ J. M{\o}rk }}\ and\ \citenamefont
  {Lippi}(2018)}]{doi:10.1063/1.5022958}%
  \BibitemOpen
  \bibfield  {author} {\bibinfo {author} {\bibnamefont {{ J. M{\o}rk }}}\ and\
  \bibinfo {author} {\bibfnamefont {G.~L.}\ \bibnamefont {Lippi}},\ }\bibfield
  {title} {\bibinfo {title} {Rate equation description of quantum noise in
  nanolasers with few emitters},\ }\href@noop {} {\bibfield  {journal}
  {\bibinfo  {journal} {Appl. Phys. Lett.}\ }\textbf {\bibinfo {volume}
  {112}},\ \bibinfo {pages} {141103} (\bibinfo {year} {2018})}\BibitemShut
  {NoStop}%
\bibitem [{\citenamefont {nan Zhang}\ \emph {et~al.}(2015)\citenamefont {nan
  Zhang}, \citenamefont {Zhao},\ and\ \citenamefont {qing Lv}}]{ZHANG2015374}%
  \BibitemOpen
  \bibfield  {author} {\bibinfo {author} {\bibfnamefont {Y.}~\bibnamefont {nan
  Zhang}}, \bibinfo {author} {\bibfnamefont {Y.}~\bibnamefont {Zhao}},\ and\
  \bibinfo {author} {\bibfnamefont {R.}~\bibnamefont {qing Lv}},\ }\bibfield
  {title} {\bibinfo {title} {A review for optical sensors based on photonic
  crystal cavities},\ }\href
  {https://doi.org/https://doi.org/10.1016/j.sna.2015.07.025} {\bibfield
  {journal} {\bibinfo  {journal} {Sensors and Actuators A: Physical}\ }\textbf
  {\bibinfo {volume} {233}},\ \bibinfo {pages} {374} (\bibinfo {year}
  {2015})}\BibitemShut {NoStop}%
\bibitem [{\citenamefont {Horsthemke}\ and\ \citenamefont
  {Lefever}(1984)}]{Horsthemke1984}%
  \BibitemOpen
  \bibfield  {author} {\bibinfo {author} {\bibfnamefont {W.}~\bibnamefont
  {Horsthemke}}\ and\ \bibinfo {author} {\bibfnamefont {R.}~\bibnamefont
  {Lefever}},\ }\href@noop {} {\emph {\bibinfo {title} {Noise-Induced
  Transitions Theory and Applications in Physics, Chemistry, and Biology}}}\
  (\bibinfo  {publisher} {Springer London, Limited},\ \bibinfo {year} {1984})\
  p.\ \bibinfo {pages} {322}\BibitemShut {NoStop}%
\bibitem [{\citenamefont {Mandel}(1995)}]{Mandel1995}%
  \BibitemOpen
  \bibfield  {author} {\bibinfo {author} {\bibfnamefont {E.}~\bibnamefont
  {Mandel}, \bibfnamefont {L.~Wolf}},\ }\href@noop {} {\emph {\bibinfo {title}
  {Optical coherence and quantum optics}}}\ (\bibinfo  {publisher} {Cambridge
  University Press},\ \bibinfo {year} {1995})\ p.\ \bibinfo {pages}
  {1166}\BibitemShut {NoStop}%
\bibitem [{\citenamefont {Jahnke}\ \emph {et~al.}(2016)\citenamefont {Jahnke},
  \citenamefont {Gies}, \citenamefont {A{\ss}mann}, \citenamefont {Bayer},
  \citenamefont {Leymann}, \citenamefont {Foerster}, \citenamefont {Wiersig},
  \citenamefont {Schneider}, \citenamefont {Kamp},\ and\ \citenamefont
  {H{\"o}fling}}]{Jahnke}%
  \BibitemOpen
  \bibfield  {author} {\bibinfo {author} {\bibfnamefont {F.}~\bibnamefont
  {Jahnke}}, \bibinfo {author} {\bibfnamefont {C.}~\bibnamefont {Gies}},
  \bibinfo {author} {\bibfnamefont {M.}~\bibnamefont {A{\ss}mann}}, \bibinfo
  {author} {\bibfnamefont {M.}~\bibnamefont {Bayer}}, \bibinfo {author}
  {\bibfnamefont {H.~A.~M.}\ \bibnamefont {Leymann}}, \bibinfo {author}
  {\bibfnamefont {A.}~\bibnamefont {Foerster}}, \bibinfo {author}
  {\bibfnamefont {J.}~\bibnamefont {Wiersig}}, \bibinfo {author} {\bibfnamefont
  {C.}~\bibnamefont {Schneider}}, \bibinfo {author} {\bibfnamefont
  {M.}~\bibnamefont {Kamp}},\ and\ \bibinfo {author} {\bibfnamefont
  {S.}~\bibnamefont {H{\"o}fling}},\ }\bibfield  {title} {\bibinfo {title}
  {Giant photon bunching, superradiant pulse emission and excitation trapping
  in quantum-dot nanolasers},\ }\href {https://doi.org/10.1038/ncomms11540}
  {\bibfield  {journal} {\bibinfo  {journal} {Nature Commun.}\ }\textbf
  {\bibinfo {volume} {7}},\ \bibinfo {pages} {11540} (\bibinfo {year}
  {2016})}\BibitemShut {NoStop}%
\bibitem [{\citenamefont {Gies}\ \emph {et~al.}(2007)\citenamefont {Gies},
  \citenamefont {Wiersig}, \citenamefont {Lorke},\ and\ \citenamefont
  {Jahnke}}]{PhysRevA.75.013803}%
  \BibitemOpen
  \bibfield  {author} {\bibinfo {author} {\bibfnamefont {C.}~\bibnamefont
  {Gies}}, \bibinfo {author} {\bibfnamefont {J.}~\bibnamefont {Wiersig}},
  \bibinfo {author} {\bibfnamefont {M.}~\bibnamefont {Lorke}},\ and\ \bibinfo
  {author} {\bibfnamefont {F.}~\bibnamefont {Jahnke}},\ }\bibfield  {title}
  {\bibinfo {title} {Semiconductor model for quantum-dot-based microcavity
  lasers},\ }\href {https://doi.org/10.1103/PhysRevA.75.013803} {\bibfield
  {journal} {\bibinfo  {journal} {Phys. Rev. A}\ }\textbf {\bibinfo {volume}
  {75}},\ \bibinfo {pages} {013803} (\bibinfo {year} {2007})}\BibitemShut
  {NoStop}%
\bibitem [{\citenamefont {Wu}\ and\ \citenamefont {Lin}(1984)}]{Wu_1984}%
  \BibitemOpen
  \bibfield  {author} {\bibinfo {author} {\bibfnamefont {W.}~\bibnamefont
  {Wu}}\ and\ \bibinfo {author} {\bibfnamefont {Y.}~\bibnamefont {Lin}},\
  }\bibfield  {title} {\bibinfo {title} {Cumulant-neglect closure for
  non-linear oscillators under random parametric and external excitations},\
  }\href {https://doi.org/10.1016/0020-7462(84)90063-5} {\bibfield  {journal}
  {\bibinfo  {journal} {International Journal of Non-Linear Mechanics}\
  }\textbf {\bibinfo {volume} {19}},\ \bibinfo {pages} {349} (\bibinfo {year}
  {1984})}\BibitemShut {NoStop}%
\bibitem [{\citenamefont {Sun}\ and\ \citenamefont
  {Hsu}(1987)}]{10.1115/1.3173083}%
  \BibitemOpen
  \bibfield  {author} {\bibinfo {author} {\bibfnamefont {J.-Q.}\ \bibnamefont
  {Sun}}\ and\ \bibinfo {author} {\bibfnamefont {C.~S.}\ \bibnamefont {Hsu}},\
  }\bibfield  {title} {\bibinfo {title} {{Cumulant-Neglect Closure Method for
  Nonlinear Systems Under Random Excitations}},\ }\href
  {https://doi.org/10.1115/1.3173083} {\bibfield  {journal} {\bibinfo
  {journal} {Journal of Applied Mechanics}\ }\textbf {\bibinfo {volume} {54}},\
  \bibinfo {pages} {649} (\bibinfo {year} {1987})}\BibitemShut {NoStop}%
\end{thebibliography}%
\end{document}